\documentclass{article}
\usepackage{lmodern}
\usepackage{arxiv}
\usepackage[utf8]{inputenc}
\usepackage[T1]{fontenc}
\usepackage{amsmath,amssymb}
\usepackage{booktabs}
\usepackage{graphicx}
\usepackage[table]{xcolor}
\usepackage{array}
\usepackage{tabularx}
\usepackage{listings}
\usepackage{microtype}
\usepackage{caption}
\usepackage{float}
\usepackage[affil-it]{authblk}
\usepackage[numbers,sort&compress]{natbib}
\usepackage{siunitx}
\usepackage{hyperref}
\usepackage{soul}
\usepackage{placeins}
\usepackage{etoolbox}

\renewcommand{\shorttitle}{Symmetrix-XL}

\hypersetup{
colorlinks=true,
linkcolor=blue,
citecolor=blue,
urlcolor=blue,
pdftitle={Train for Accuracy, Execute at Scale: Architecture-Preserving Inference for Equivariant Atomistic Foundation Models},
pdfsubject={High-performance inference for equivariant foundation models},
pdfauthor={Lei Fu, Zihui Feng, Yongheng Li, Hongwei Du, Xin He, Junyi Wu, Kejie Bao, Yueyu Zhang, Ziheng Lu, Bonan Zhu},
pdfkeywords={equivariant foundation models, MACE, streamed-edge execution, model specialization, molecular dynamics}
}

\newcommand{\result}[1]{#1}
\newcommand{\resultcell}[1]{#1}
\newcolumntype{Y}{>{\raggedright\arraybackslash}X}

\DeclareSIUnit\angstrom{\text{\AA}}
\DeclareUnicodeCharacter{207A}{\ensuremath{^{+}}}
\DeclareUnicodeCharacter{207B}{\ensuremath{^{-}}}
\DeclareUnicodeCharacter{2074}{\ensuremath{^{4}}}
\DeclareUnicodeCharacter{2077}{\ensuremath{^{7}}}
\DeclareUnicodeCharacter{2085}{\ensuremath{_{5}}}
\DeclareUnicodeCharacter{2086}{\ensuremath{_{6}}}
\DeclareUnicodeCharacter{03BC}{\ensuremath{\mu}}
\DeclareUnicodeCharacter{0394}{\ensuremath{\Delta}}
\DeclareUnicodeCharacter{00D7}{\ensuremath{\times}}

\providecommand{\chem}[1]{\ensuremath{\mathrm{#1}}}
\providecommand{\NOtwo}{\chem{NO_2}}
\providecommand{\HONO}{\chem{HONO}}

\providecommand{\CS}{\chem{C}\text{--}\chem{S}}
\providecommand{\CH}{\chem{C}\text{--}\chem{H}}

\graphicspath{{figures/}}
\BeforeBeginEnvironment{abstract}{\vspace{-2em}}

\title{Train for Accuracy, Execute at Scale:\\
Architecture-Preserving Inference for Equivariant Atomistic Foundation Models}

\author[1,2,3]{Lei Fu}
\author[2,3]{Zihui Feng}
\author[2,3]{Yongheng Li}
\author[2,3]{Hongwei Du}
\author[2,3]{Xin He}
\author[4]{Junyi Wu}
\author[4]{Kejie Bao}
\author[4]{Yueyu Zhang}
\author[5]{Zeyu Deng}
\author[2,3,4,*]{Ziheng Lu}
\author[1,6,\textdagger]{Bonan Zhu}

\affil[1]{School of Aerospace Engineering, Beijing Institute of Technology, 5 Zhongguancunnan Street, Beijing, 100081, China}
\affil[2]{Zhongguancun Academy, Beijing, 100094, China}
\affil[3]{Zhongguancun Institute of Artificial Intelligence, Beijing, 100094, China}
\affil[4]{Kairos Materials}
\affil[5]{National University of Singapore, 9 Engineering Drive 1, Singapore, 117575 Singapore}
\affil[6]{State Key Laboratory of Environment Characteristics and Effects for Near-space, Beijing, 100081, China}

\date{}

\begin{document}
\raggedbottom

\setcounter{topnumber}{5}
\setcounter{bottomnumber}{5}
\setcounter{totalnumber}{10}

\renewcommand{\topfraction}{0.95}
\renewcommand{\bottomfraction}{0.95}
\renewcommand{\textfraction}{0.05}
\renewcommand{\floatpagefraction}{0.8}

\maketitle

\begingroup
\renewcommand{\thefootnote}{\fnsymbol{footnote}}
\makeatletter
\renewcommand\@makefntext[1]{%
  \noindent
  \hb@xt@0.2em{\hss\@makefnmark}#1%
}
\makeatother

\footnotetext[1]{Corresponding: \href{mailto:zluag@connect.ust.hk}{zluag@connect.ust.hk}}
\footnotetext[2]{Corresponding: \href{mailto:bzhu@bit.edu.cn}{bzhu@bit.edu.cn}}
\endgroup

\begin{abstract}

Equivariant atomistic foundation models provide broadly transferable interatomic potentials trained against quantum-mechanical reference data, but their repeated execution at simulation scale remains computationally and memory intensive. We present Symmetrix-XL, an inference engine that scales pretrained MACE checkpoints without retraining, distillation, or modification of their learned weights. It combines streamed-edge execution to avoid graph-wide materialization of expanded edge intermediates, model-specialized code generation to compile checkpoint-specific operators, and tiled execution to reuse a bounded device-memory workspace. On complete LAMMPS-step benchmarks, Symmetrix-XL reduces inference time by 3.1-5.0 times relative to ML-IAP + cuEquivariance across tested A100 and RTX 5090 workloads. On a single A100 80 GB GPU, the tested MACE-OMAT-0 capacity boundary increases from 24,565 atoms to 11.24 million atoms. This substantially lowers the hardware threshold for simulations that would otherwise require spatial decomposition across many GPUs and compute nodes. The same backend weak-scales to 703 million atoms on 64 A800 GPUs at 93.8 percent efficiency. Energy, force, stress, molecular-dynamics stability, and Matbench Discovery evaluations reproduce reference MACE behavior within measured tolerances. Case studies spanning solid-state, interfacial, and reactive systems demonstrate the increased capacity in realistic workflows. These results establish post-training execution as a scaling axis complementary to model redesign, compression, and distributed scale-out, and show that the practical accuracy-cost frontier depends on both model architecture and inference execution.

\end{abstract}

\vspace{1.0em}

\vspace{1.0em}
\section{Introduction}
\label{sec:introduction}

Atomistic foundation models, also known as universal machine-learning interatomic potentials, are increasingly trained at large scale and then reused across broad chemical and structural spaces \cite{matbench,mace_mp0,omat24,uma, Prophet, ufm_takamoto2022_pfp,
ufm_chen2022_m3gnet,
ufm_deng2023_chgnet,
ufm_choudhary2023_alignnff,
ufm_batatia2025_macemp,
ufm_yang2024_mattersim,
ufm_park2024_sevennet,
ufm_kim2025_sevennetmf,
ufm_barrosluque2026_omat24,
ufm_neumann2024_orb,
ufm_rhodes2025_orbv3,
ufm_zhang2024_dpa2,
ufm_zhang2025_dpa3,
ufm_lysogorskiy2026_grace,
ufm_fu2025_esen,
ufm_wood2025_uma,
ufm_mazitov2025_petmad,
ufm_koker2025_nequix,
ufm_kavanagh2026_nequipallegro,
ufm_liang2025_nep89}. As with large language models, this creates a natural separation between training and inference: the system used to train a flexible model family need not be the system used to repeatedly execute one fixed checkpoint. This distinction is particularly important for atomistic simulations, where the same potential may be evaluated millions of times in a molecular-dynamics trajectory. Inference speed determines the accessible time scale, while the memory required by the neighbor graph and intermediate tensors limits the largest system that can be simulated.

One route to lower cost is to design the deployed model for efficient
simulation. NEP combines compact atom-centered descriptors with a shallow
neural network and an optimized GPU implementation in GPUMD
\cite{ufm_fan2021_nep,ufm_fan2022_gpumd}, and NEP89 extends this approach
to a foundation model spanning 89 elements
\cite{ufm_liang2025_nep89}. Nequix combines a compact four-layer
NequIP-derived model with custom fused kernels \cite{nequix}, UMA uses
eSCN/\(\mathrm{SO}(2)\) message-passing blocks and a mixture of linear
experts \cite{uma}, and E2Former uses Wigner-\(6j\) recoupling to shift
tensor-product work from edges to nodes \cite{e2former}. DPA4C goes
further by removing message passing of learned atomic states, making its
interaction locality effectively equivalent to that of a local-descriptor
model such as Allegro \cite{dpa4c,allegro}. Knowledge distillation provides
another route, with a foundation model supplying labels for a smaller
deployment model, as demonstrated by SevenNet-Nano \cite{sevennet_nano}.
These approaches combine architectural choices, training strategies, and
implementation optimizations to reduce deployment cost. When the architecture
or learned parameters are changed, however, the resulting potential is a
different model and requires separate validation.

\begin{figure*}[t]
    \centering
    \includegraphics[width=\textwidth]{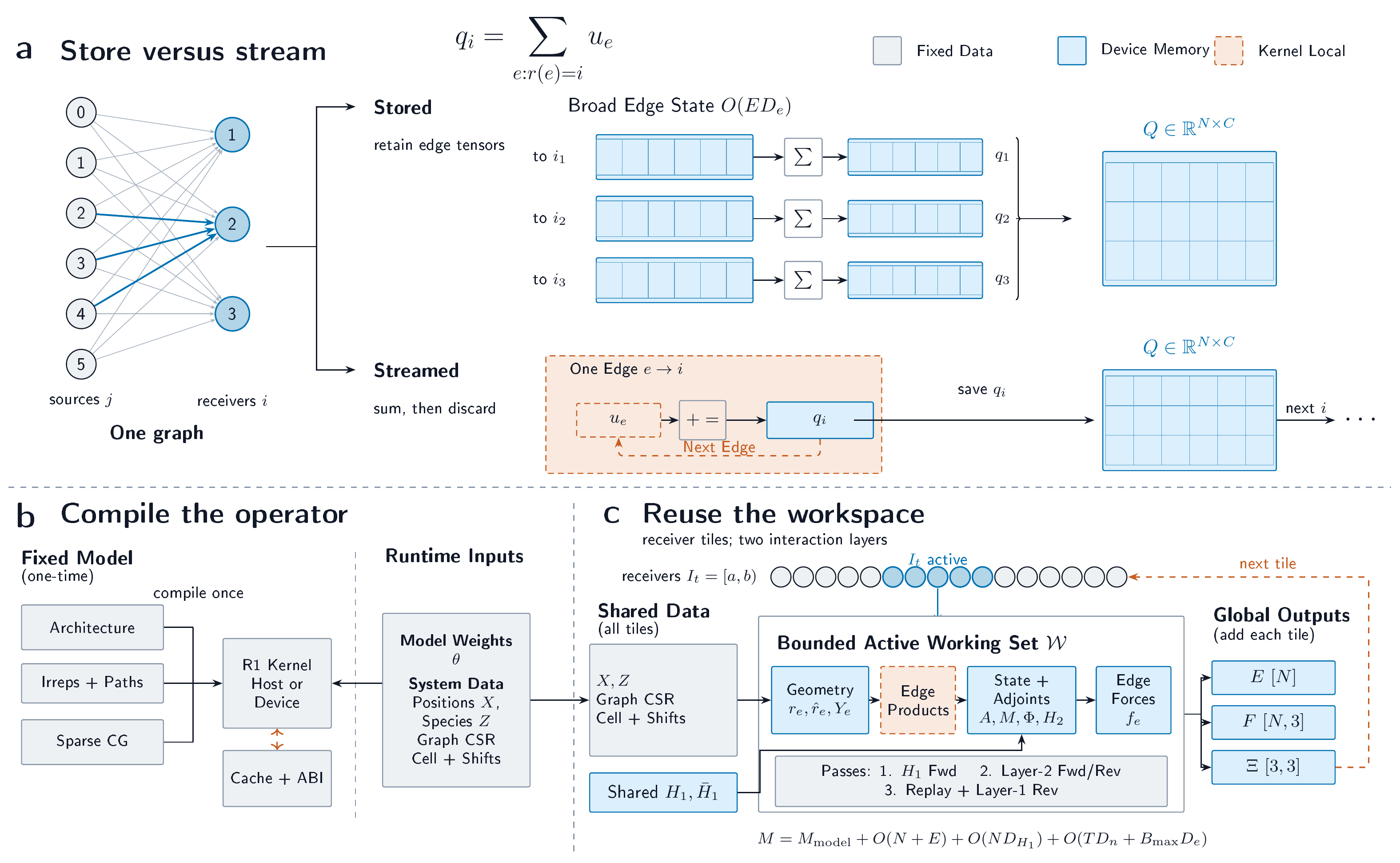}
    \caption{
        Overview of the Symmetrix-XL execution strategy.
        \textbf{(a)} Streamed-edge execution avoids retaining large
        graph-wide radial and tensor-product edge states by generating each
        contribution near its point of use, accumulating it into the
        corresponding receiver, and discarding kernel-local products.
        \textbf{(b)} Model-specialized execution lowers the checkpoint's fixed
        sparse coupling structure, coefficients, and tensor layouts into a
        compiled and cached contraction program. Learned model parameters and
        per-system graph and geometry data remain runtime inputs.
        \textbf{(c)} For capacity-limited calculations, optional tiled
        execution partitions receivers and their incident edges into bounded
        tiles and reuses a tile-local workspace across three sweeps.
        The inter-layer state $H_1$ and its adjoint $\bar{H}_1$ remain
        graph-wide, while larger temporary intermediates are bounded by the
        active tile. Persistent energy, force, virial, and inter-layer updates
        are committed before the workspace is reused.
    }
    \label{fig:symmetrix_xl_overview}
\end{figure*}

A different question is how far an already-trained foundation model can
be scaled without replacing it. Tan, Witt, and co-workers demonstrated that
specialized execution can substantially accelerate deep equivariant
interatomic potentials without redesigning the underlying model
\cite{nequip_deploy}. Recent NequIP and Allegro work has likewise achieved
substantial acceleration through compilation, optimized tensor-product
kernels, and scalable deployment \cite{kavanagh2026_nequip_foundation}.
Related kernel generators fuse sparse Clebsch--Gordan tensor products with
graph aggregation and exploit their fixed sparse structure, reducing
intermediate storage and execution overhead
\cite{openequivariance,flashtp}. The chemtrain-deploy work connects
compiled JAX potentials to domain-decomposed LAMMPS simulations
\cite{chemtrain_deploy}, while optimized simulation interfaces and libraries
such as cuEquivariance further reduce the cost of executing existing model
programs. These results show that substantial gains remain available after
training. The question addressed here is whether such execution-level
optimization can change the practical simulation scale of an atomistic
foundation model while retaining its learned weights and architectural
contract.

The contribution of this work is therefore not a new model architecture or training framework, but a set of execution improvements for an already trained model.
First, streamed-edge execution forms expanded edge contributions only when needed, accumulates them into receiver-owned state, and recomputes required factors during the analytic reverse pass instead of retaining broad edge tensors.
Second, model-specialized execution emits compiled CPU or GPU operations from the fixed Clebsch--Gordan paths, coefficients, tensor layouts, and channel dimensions of a trained checkpoint.
Third, tiled execution partitions receivers and their incident edges into bounded work units and reuses a fixed temporary workspace to extend single-accelerator capacity.

We implement and evaluate these techniques on MACE as a representative multi-layer equivariant foundation model.
The implementation, Symmetrix-XL, builds on the original Symmetrix inference backend, which already reorganized important parts of the equivariant contraction program but retained a materialized execution path for expanded graph-wide intermediates \cite{symmetrix_original}.
The checkpoint's learned weights, irreducible representations, channel widths, product basis, readouts, and number of interactions are retained; the work changes the execution program rather than the trained architecture.
This does not imply bitwise identity: operation ordering and radial spline projection can change the numerical realization, so equivalence is established through explicit numerical and downstream validation.
These techniques may also apply to other equivariant graph-neural-network foundation models, subject to architecture-specific implementation and validation.

We refer to the resulting increase in the system size accommodated by each GPU as \emph{single-accelerator scale-up}.
Large atomistic simulations can also be assembled by spatial decomposition across many GPUs, but that distributed scale-out does not remove the per-rank memory constraint imposed by the model and its intermediates.
Prior work has established scalable spatial-decomposition execution for multilayer equivariant message-passing potentials \cite{park2024sevennet,kavanagh2026_nequip_foundation}.
Increasing per-GPU capacity is therefore complementary to MPI scaling: it permits larger local domains, can reduce the minimum number of GPUs required for a given system, and provides a larger building block for distributed simulations.

\section{Execution pattern for equivariant message passing}
\label{sec:mace}

A broad class of equivariant atomistic models evaluates a sparse neighbor graph by forming edge-conditioned equivariant contributions and reducing them into receiver-owned node states.
The precise representations and update rules differ between architectures, but this execution pattern creates a common systems problem: large edge-local intermediates can be produced before being reduced to substantially smaller node-local state.
We use MACE as the concrete instantiation of this pattern throughout this work \cite{mace2022}.

MACE represents a structure as a directed radius graph \cite{mace2022}.
For a directed edge \(e=(j\rightarrow i)\), the model evaluates invariant radial features from the distance \(d_e\) and real spherical harmonics \(Y_{\ell m}(\hat{\mathbf r}_e)\) from the edge direction.
Here atom \(j\) is the \emph{source} (the neighboring atom that supplies features), atom \(i\) is the \emph{receiver} (the central atom whose state is updated), and we write \(s(e)=j\) and \(r(e)=i\).
The node state at interaction \(t\) is a direct sum of irreducible representations,
\[
h_i^{(t)}=\bigoplus_{\ell,p}h_i^{(t,\ell,p)},
\qquad
h_i^{(t,\ell,p)}\in\mathbb R^{C_{\ell p}\times(2\ell+1)}.
\]
An interaction first applies a learned equivariant map to the source state,
\[
x_j^{(t)}=L_{\mathrm{up},t}h_j^{(t)},
\]
then forms edge-conditioned tensor products and sums them at each receiver,
\[
u_{e,\pi c m_o}^{(t)}
=
w_{e,\pi c}^{(t)}
\sum_{m_i,m_y}
C^{\pi}_{m_i m_y m_o}
x_{s(e),c m_i}^{(t)}
Y_{\ell_y m_y}(\hat{\mathbf r}_e),
\qquad
q_i^{(t)}=\sum_{e:r(e)=i}u_e^{(t)}.
\]
Here \(C^{\pi}\) is a sparse Clebsch--Gordan coupling path, \(w_{e,\pi c}^{(t)}\) is produced by the radial network, and \(q_i^{(t)}\) denotes the direct sum of the path-resolved receiver outputs.
A species-conditioned symmetric product basis then raises the effective body order without explicitly enumerating neighbor tuples.
Readouts from successive interactions sum to the total energy, and forces and stress are derivatives of that scalar with respect to coordinates and cell.

MACE provides a particularly useful instantiation of this execution pattern for three reasons.
Its two main interactions repeatedly apply a fixed sparse equivariant program to a graph whose size changes at runtime.
Its high-body-order product basis increases expressivity per interaction, allowing common foundation checkpoints to use comparatively few message-passing layers (typically two).
Its radial functions depend only on distance and species and can therefore be projected once into compact spline tables, avoiding recomputation during repeated inference.

The reference \texttt{mace-torch} implementation expresses this computation as a PyTorch program built from e3nn primitives \cite{pytorch,e3nn,mace_software}.
For energy, force, and stress evaluation, reverse-mode automatic differentiation retains the intermediates required by the backward pass.
This creates substantial graph-proportional memory traffic and storage.
Operator libraries such as cuEquivariance, OpenEquivariance, and FlashTP accelerate sparse Clebsch--Gordan contractions and may fuse them with graph aggregation, but the surrounding framework still manages the wider model program and its graph-wide state \cite{cuequivariance,openequivariance,flashtp}.
We use PyTorch~+~cuEquivariance as the optimized PyTorch baseline for the matched ASE benchmark.
Complete calculator calls use the Atomic Simulation Environment \cite{ase}.
For large-scale molecular-dynamics simulations, we choose ML-IAP-Kokkos~+~cuEquivariance as an optimized MACE baseline.
This stack combines LAMMPS's scalable ML-IAP/Kokkos integration with NVIDIA's vendor-optimized binary cuEquivariance CUDA kernels, whose CUDA-major-specific acceleration packages are distributed as binary wheels \cite{lammps2022,lammps_kokkos,cuequivariance,nvidia_mliap_mace}.

\section{Streamed-edge execution}
\label{sec:streaming}

The first optimization concerns how long edge-local intermediates remain live.
When edge contributions are ultimately reduced into receiver-owned state, the expanded quantities need not necessarily remain materialized over the full graph.
For the equivariant tensor-product interaction above, we build on the streamed formulation of Choro\v{s}ajev and B\'{e}ny \cite{sobek}.
Write the expanded edge weight as a projection of a compact radial embedding \(\varphi_{eq}\),
\[
W_{e,\pi a}=\sum_q\varphi_{eq}B_{q,\pi a},
\]
and the receiver message as
\[
m_i=\sum_{e:r(e)=i}\sum_{\pi,a}W_{e,\pi a}T_{\pi a}(h_{s(e)},Y_e).
\]
Here \(q\) indexes the compact radial embedding, \(\pi\) identifies an admitted tensor-product path, \(a\) indexes its expanded radial/path component, \(B\) is the learned projection from the compact embedding, and \(T_{\pi a}\) denotes the corresponding angular/source-feature tensor-product contribution.
The expanded weights \(W\), edge messages, and their adjoints are scheduling intermediates rather than model outputs.
Substitution permits each expanded contribution to be formed, accumulated into its receiver, and discarded without changing the algebraic operator.
The reverse pass applies the transpose of the same reassociation and recomputes the required radial and angular factors under receiver-, source-, or edge-owned work schedules.

Earlier Symmetrix execution already reorganized parts of the contraction algebra and used a materialized schedule for these expanded graph-wide intermediates.
Relative to that path, Symmetrix-XL changes their storage lifetime: expanded quantities are consumed during aggregation and, where needed, recomputed during the analytic reverse pass.
For \(N\) atoms and \(E\) directed edges, a materialized reverse pass can retain terms proportional to
\[
\mathcal O(ECP)+\mathcal O(EW_R)+\mathcal O(NCG),
\]
where \(C\) is the channel count, \(P\) the tensor-product path count, \(W_R\) the radial width, and \(G\) the product-basis state per channel.
For a model-aware description, let \(K_{\mathrm{geom}}\) denote the fixed number of stored topology, geometry, and harmonic components per edge, \(K_{\mathrm{node}}\) the checkpoint-dependent width of the persistent node-boundary states in units of channels, and \(K_{\mathrm{work}}\) the bounded kernel-local workspace selected by the launch plan.
The streamed schedule then reduces persistent graph-dependent storage to
\[
\mathcal O(EK_{\mathrm{geom}})_{\mathrm{graph/geometry}}
+\mathcal O(NC K_{\mathrm{node}})_{\mathrm{node\ boundaries}}
+\mathcal O(K_{\mathrm{work}})_{\mathrm{bounded\ work}}.
\]
For a fixed checkpoint and launch plan, \(K_{\mathrm{geom}}\), \(K_{\mathrm{node}}\), and \(K_{\mathrm{work}}\) are constants with respect to \(N\) and \(E\), recovering the shorthand \(\mathcal O(E)+\mathcal O(NC)+\mathcal O(1)\).
The last term denotes model- and launch-bounded temporaries held in registers, cache, or backend team/shared memory while the expanded edge hot path executes; it does not include the graph and node arrays that remain in device memory.
This does not remove all edge-dependent storage, and capacity still depends on cutoff, neighbor density, precision, and the retained execution plan.
It instead removes the broad expanded arrays whose width grows with channels and coupling paths.

We apply this lifetime policy to both the forward and analytic reverse schedules.
Expanded second-interaction radial and coupling values are evaluated while forming the receiver sum \(m_i\) above; only after that sum is complete does a learned linear map project the node-sized representation into the next hidden state.
Applying this dense projection once per receiver avoids storing a projected representation for every edge.
Reverse mode first applies the transpose of that node projection and then recomputes local factors in generated source- and edge-owned kernels.
The runtime selects a qualified retention or recomputation plan from model and graph metadata before measurement begins.

\section{Model-specialized execution}
\label{sec:specialization}


A training framework must interpret many possible architectures, whereas inference repeatedly executes one fixed operator.
This creates a general opportunity for specialization after training.
In the MACE implementation studied here, the nonzero Clebsch--Gordan terms, their coefficients, tensor layouts, and channel dimensions are known before a simulation begins.
Symmetrix-XL writes these fixed operations into generated source code, exposing regular loops and constant coupling data to the CPU or GPU compiler.
Atom counts, neighbor graphs, and learned weights remain runtime inputs, so the generated program is specialized to the model architecture rather than to one simulation cell.

Host artifacts are compiled as C++20 shared libraries, while CUDA and HIP artifacts use NVRTC and hipRTC, respectively \cite{nvrtc,hiprtc}.
Artifacts are cached and validated against the model contract; production execution fails rather than silently falling back when a required artifact cannot be loaded.
They can be prepared before a cluster or LAMMPS job, so compilation is excluded from steady-state simulation timing.

The generated contraction is part of a larger native evaluator rather than a generated evaluator in its entirety.
Kokkos owns graph data, geometry, node stages, reverse scheduling, and reductions, while SpheriCart supplies spherical harmonics \cite{kokkos,sphericart}.

\section{Tiled execution}
\label{sec:tiled}

Symmetrix-XL provides two streamed-edge execution modes: standard streaming and tiled streaming.
Standard streaming processes the graph directly without a persistent fixed workspace, whereas tiled streaming partitions the same streamed computation into bounded tiles and reuses a fixed workspace in device memory.
For graphs that exceed the capacity of standard streaming, tiled streaming can therefore extend capacity while retaining kernel-local edge streaming within each tile; it is called \textit{fixed-workspace} execution internally.
Receivers and their incident edges are partitioned into bounded tiles, and the same temporary workspace is reused across three sweeps: the first-interaction forward pass, the second-interaction forward and reverse passes, and a replay followed by the first-interaction reverse pass.
For a two-interaction MACE model, the intermediate per-node hidden features \(H_1\) and their adjoints remain graph-wide because they connect the two interaction layers, whereas the larger \(A\), \(M\), \(\Phi\), geometry, harmonic, and directed-force workspaces are bounded by the largest active tile.
Tiled execution therefore reduces the dominant temporary workspace from graph-sized to tile-sized storage while retaining the checkpoint's architecture and learned weights, at the cost of additional scheduling, DRAM traffic, and recomputation.
Because this trade-off targets capacity-limited calculations, the execution planner considers the tiled plan only when it is explicitly enabled.

\section{Benchmark protocol}
\label{sec:protocol}

The remainder of the paper evaluates the MACE instantiation of these execution ideas.
The benchmark data were collected during development of the reported execution paths.
The historical original-Symmetrix control uses commit \texttt{e762729} from the original repository \cite{symmetrix_original}.
Unless otherwise stated, the primary performance and capacity benchmarks use the MACE-OMAT-0 medium checkpoint.

The distributed LAMMPS scaling results use A800 GPUs and the standard-streaming and tiled-streaming modes.
We use ASE for complete calculator-call measurements representative of common single-device workflows, including molecular dynamics, geometry optimization, and phonon calculations.
We use LAMMPS for sustained molecular dynamics and spatially decomposed multi-GPU execution.
Performance is reported primarily in \(\mu\mathrm{s}/\mathrm{atom}\), with complete step or call time as secondary context.

The primary reference workloads use the same checkpoint, coordinates, precision, and requested properties for Symmetrix-XL, PyTorch/e3nn, and PyTorch~+~cuEquivariance.
A \emph{complete ASE calculator call} includes Python calculator dispatch, graph construction or retained-neighbor input update, native or framework evaluation, and assembly of the returned energy, forces, and stress.
The ASE values therefore include graph-build time and are end-to-end deployment measurements, not isolated model-forward timings.

The matched LAMMPS throughput comparisons use float32 on either an A100-SXM4 80~GB or an RTX 5090 32~GB GPU, with a 0.5~\AA{} skin, five warm-up steps, 20 measured steps, and three fresh-process trials per implementation.
The primary statistic is the median LAMMPS Loop time in \(\mu\mathrm{s}/\mathrm{atom}/\mathrm{step}\).
Coordinates are static during these timing blocks, and all accepted trials report zero neighbor-list rebuilds and zero dangerous builds.
Capacity is bracketed by the largest successful cubic SrTiO\(_3\) repeat and the adjacent failed repeat.

The scaling measurements use moving NVT dynamics with a neighbor-list skin.
The record reports the skin, effective cutoff, actual rebuild count, and directed-edge count so topology reuse is explicit.
Single-GPU capacity is the largest graph for which the stated call or repeated-step protocol completes without an out-of-memory failure.
An evaluator-call capacity result is not described as sustained molecular dynamics unless it passes the repeated 20-step protocol.

We separately evaluated whether the optimized execution path preserves the predictions and downstream results of the original MACE implementation.
This qualification used the MACE-MPA-0 medium checkpoint in float64 and compared Symmetrix-XL with \texttt{mace-torch} on identical atomic configurations.
We then repeated the Matbench Discovery WBM relaxation \cite{matbench}, thermal-conductivity protocol \cite{pota_thermal}, published MACE-MPA-0 diatomic evaluation \cite{mace_mp0}, and DynaMat molecular-dynamics benchmark \cite{dynamat} using their published task protocols and evaluation metrics.
Checkpoint, converted-model, dataset, executable, and generated-artifact identities are retained in the internal evidence records.

\section{Results}
\label{sec:results}

\subsection{Throughput}

\begin{figure}[!htbp]
\centering
\includegraphics[width=\textwidth]{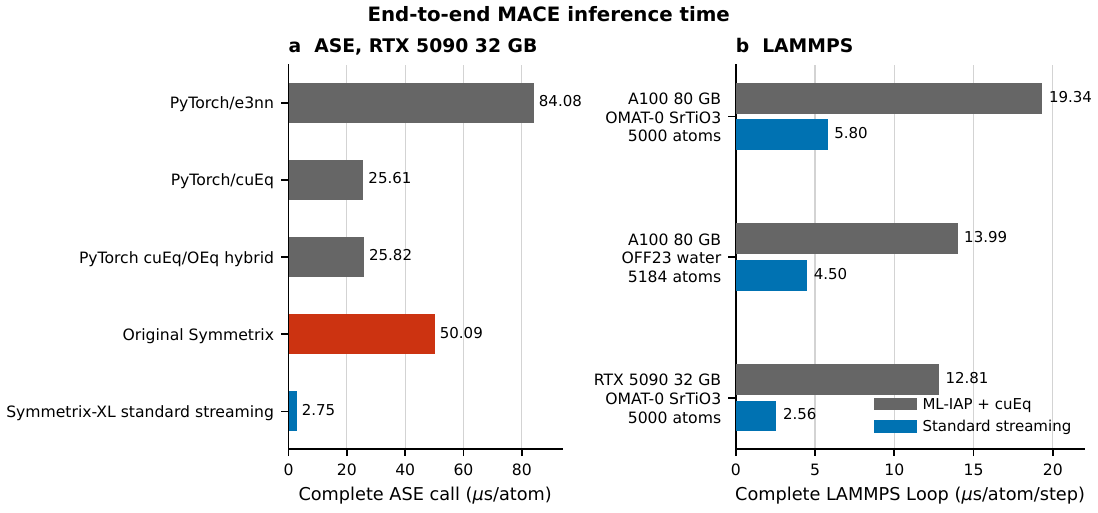}
\caption{End-to-end inference time; lower values are better. \textbf{a}, Complete ASE energy--forces--stress calls for MACE-OMAT-0 medium on 5,000-atom cubic SrTiO\(_3\) and an RTX 5090 32~GB. The boundary includes graph construction or retained-neighbor input update and result assembly. PyTorch uses an exact 6.0~\AA{} graph, whereas Symmetrix-XL uses a retained graph with a 0.5~\AA{} skin and 6.5~\AA{} effective cutoff. \textbf{b}, Complete float32 LAMMPS Loop times with the same skin policy. Values are medians of three fresh-process trials after five warm-up and 20 measured steps. OMAT-0 uses a 6.0~\AA{} model cutoff and OFF23 uses 5.0~\AA{}.}
\label{fig:headline_throughput}
\end{figure}

The ASE measurements in Figure~\ref{fig:headline_throughput} are complete call timings rather than isolated neural-network forward passes.
This end-to-end boundary intentionally includes each framework's graph construction or retained-neighbor update and calculator setup overhead, because these costs are part of deployed simulation workflows; it does not isolate raw model-inference speed.

The complete-step LAMMPS comparison shows the same ordering across both tested model--structure pairs and both GPU generations.
All accepted A100 and RTX 5090 timing runs report zero neighbor-list rebuilds and zero dangerous builds during the static measured block.

\subsection{Single-accelerator capacity}

\begin{figure}[!htbp]
\centering
\includegraphics[width=\textwidth]{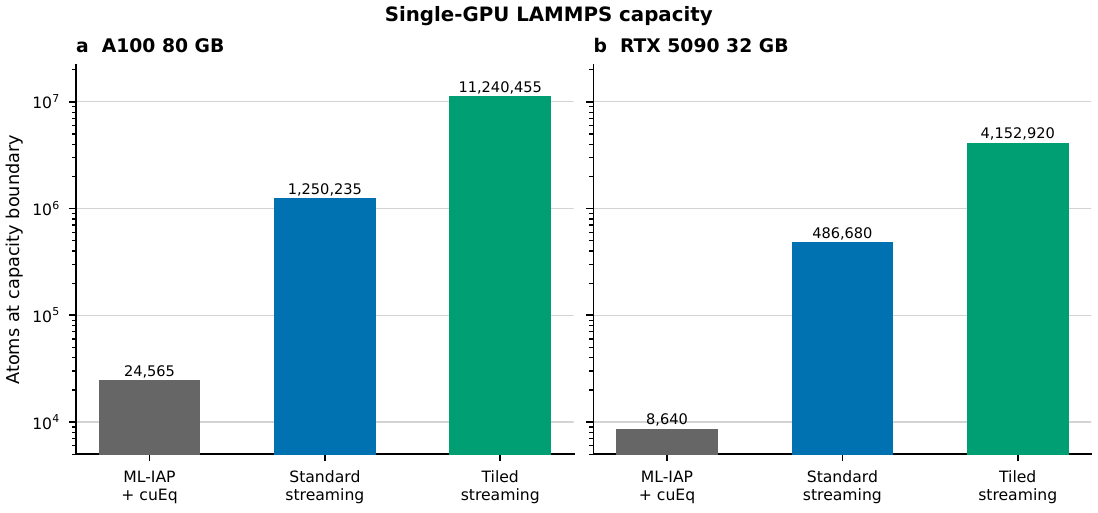}
\caption{Single-GPU SrTiO\(_3\) LAMMPS capacity for MACE-OMAT-0 medium in float32 on \textbf{a}, one A100 80~GB and \textbf{b}, one RTX 5090 32~GB. Bars show the largest successful system on the logarithmic atom-count scale. Each boundary was reproduced twice in fresh processes with a 6.0~\AA{} model cutoff and a 0.5~\AA{} neighbor skin. The A100 largest successes are 24,565, 1,250,235, and 11,240,455 atoms for ML-IAP with cuEquivariance, standard streaming, and tiled streaming, respectively; the corresponding RTX 5090 values are 8,640, 486,680, and 4,152,920 atoms.}
\label{fig:headline_capacity}
\end{figure}

\begin{table}
\centering
\small
\caption{Complete LAMMPS Loop time at each largest successful capacity boundary in Figure~\ref{fig:headline_capacity}. Values are medians of the two reproduced capacity runs. Each timing therefore corresponds to a different system size and is not a matched-size comparison of execution plans.}
\label{tab:capacity_boundary_time}
\begin{tabularx}{\linewidth}{@{}XYYYY@{}}
\toprule
Implementation & A100 atoms & A100 Loop, \(\mu\mathrm{s}/\mathrm{atom}/\mathrm{step}\) & RTX 5090 atoms & RTX 5090 Loop, \(\mu\mathrm{s}/\mathrm{atom}/\mathrm{step}\) \\
\midrule
ML-IAP~+~cuEq & \resultcell{24,565} & \resultcell{25.69} & \resultcell{8,640} & \resultcell{22.00} \\
Standard streaming & \resultcell{1,250,235} & \resultcell{6.15} & \resultcell{486,680} & \resultcell{3.06} \\
Tiled streaming & \resultcell{11,240,455} & \resultcell{6.41} & \resultcell{4,152,920} & \resultcell{3.16} \\
\bottomrule
\end{tabularx}
\end{table}

We next test the simulation capacity on a single accelerator (A100 and RTX 5090).
Standard streaming reaches its boundary at the graph-wide \(H_1\) adjoint, whereas the adjacent tiled-streaming case reaches a LAMMPS communication-buffer allocation after the Symmetrix graph plan has been constructed.
Tiled streaming uses 8,192 workspace receivers and is an explicitly enabled capacity mode rather than the default used for throughput measurements.
These capacity probes complete the initial force evaluation and one measured force evaluation and should not be interpreted as sustained 20-step molecular dynamics.

The RTX 5090 results show the same qualitative separation between standard and tiled streaming as the A100 results.
The adjacent failures occur in the standard-streaming mode's graph-wide forward state and in a LAMMPS communication buffer for tiled streaming, respectively.
At the separately determined capacity boundaries, tiled streaming increases the per-atom Loop time by approximately 4\% on each GPU relative to standard streaming while accommodating approximately nine times as many atoms.
Because the two modes are timed at different system sizes, this comparison characterizes performance at their practical capacity boundaries rather than an isolated same-size tiling overhead.

\subsection{CPU performance}

\begin{figure}[!htbp]
\centering
\includegraphics[width=\textwidth]{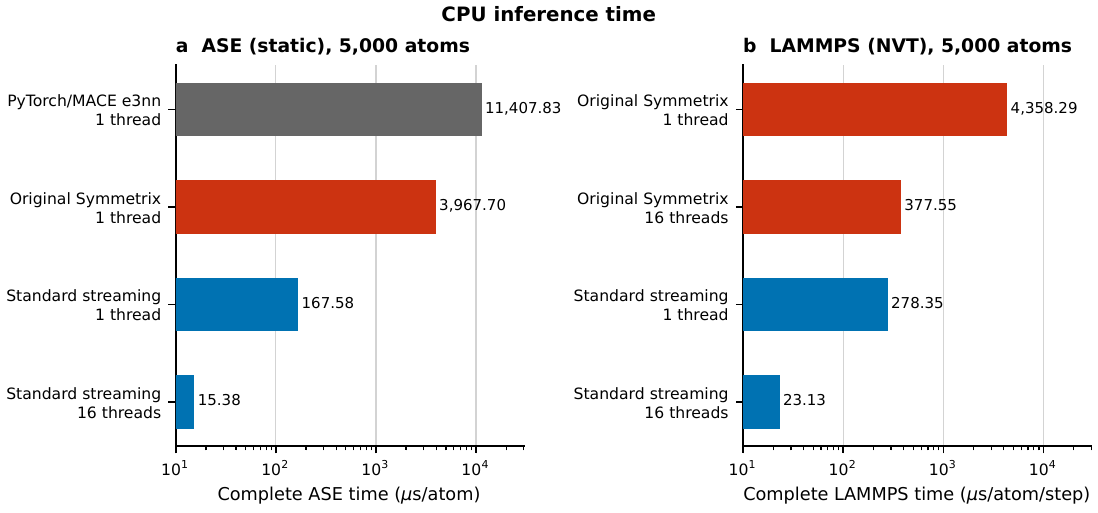}
\caption{CPU inference time for MACE-OMAT-0 medium on 5,000-atom cubic SrTiO\(_3\) and an AMD Ryzen 9 9950X3D2; lower values are better. \textbf{a}, Complete ASE energy--forces--stress calls, reported as within-process medians of ten fixed-geometry calls after three warm-up calls. The boundary includes graph construction, input preparation, model evaluation, and result assembly. PyTorch/MACE e3nn and original Symmetrix use exact 6.0~\AA{} graphs, whereas Symmetrix-XL uses a retained graph with a 0.5~\AA{} skin. \textbf{b}, Complete LAMMPS NVT Loop times, reported as medians of three fresh processes with five warm-up and 20 measured steps. The LAMMPS runs use one rank and a 6.5~\AA{} effective neighbor-list cutoff. Kokkos uses the stated number of physical cores and BLAS uses one thread throughout.}
\label{fig:cpu_inference_time}
\end{figure}

GPUs are commonly used for foundation-model inference, but CPUs remain important in workflows involving many small structures, including structure searches and pre-relaxation before density-functional-theory calculations.
Individual cells containing at most a few hundred atoms may underutilize a GPU unless multiple structures are evaluated as a batch, which adds workflow complexity.

With Symmetrix-XL, optimized CPU inference can be practical for typical MACE workloads.
The complete ASE-call comparison in Figure~\ref{fig:cpu_inference_time} shows a substantial reduction in CPU inference time.
On one thread, the specialized standard-streaming path is 23.7 times faster than original Symmetrix and 68.1 times faster than PyTorch/MACE e3nn at 5,000 atoms.
Increasing the standard-streaming path from one to 16 OpenMP threads reduces the ASE time from 167.58 to 15.38~\(\mu\mathrm{s}/\mathrm{atom}\), a 10.9-fold speedup and 68\% parallel efficiency.
No 16-thread original-Symmetrix ASE result is reported because that frontend initializes Kokkos with one worker irrespective of the requested thread count.

The complete LAMMPS boundary shows the same behavior.
Relative to original Symmetrix, the specialized standard-streaming path is 15.7 times faster on one thread and 16.3 times faster on 16 threads.
Its one-to-16-thread speedup is 12.0-fold, corresponding to 75\% parallel efficiency.
Pair evaluation accounts for nearly all Loop time in these runs, while the retained neighbor list is not rebuilt during measurement.

Model specialization is an important contributor to this CPU result.
Once the checkpoint contract is fixed, emitted host code encodes the sparse coupling paths, coefficients, layouts, and supported channel structure directly instead of interpreting them at runtime.
This fixed program exposes regular inner loops to compiler transformations such as unrolling and SIMD vectorization, including AVX2 or AVX-512 instructions when enabled by the selected host target.

We note that absolute CPU performance depends on the processor and compilation target.
These measurements use an AMD Ryzen 9 9950X3D2, and other CPU and system configurations may produce different speedups.
Appendix~\ref{app:mpi-one} reports increasing per-atom cost as the number of chemical species and associated spline tables grows.
A larger spline-table working set and reduced cache reuse are plausible contributors, although this interpretation has not been verified using hardware counters.

\subsection{Distributed scaling}

\begin{figure}[!htbp]
\centering
\includegraphics[width=\textwidth]{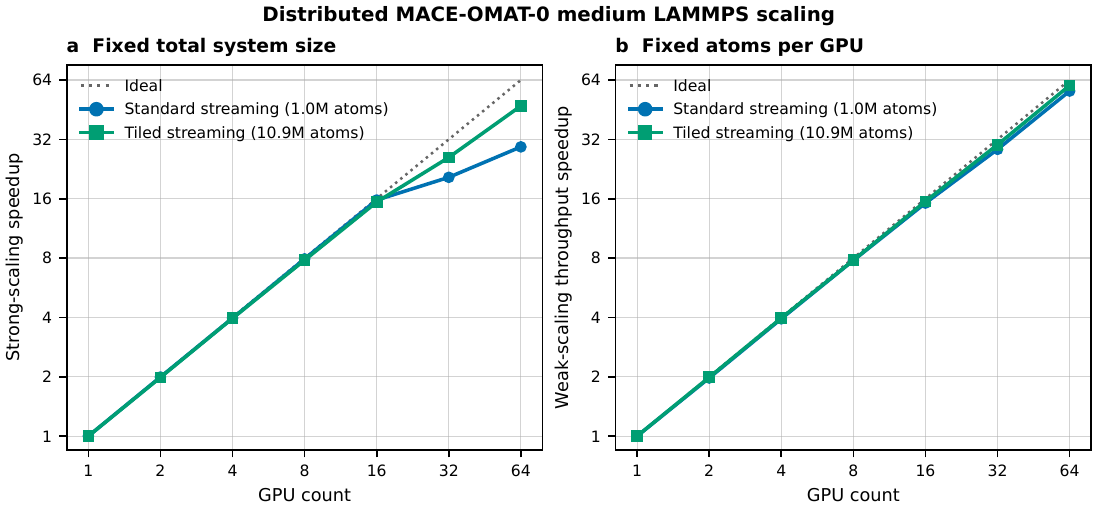}
\caption{Distributed scaling of moving-NVT MACE-OMAT-0 medium LAMMPS simulations on A800-SXM4-80GB GPUs using GPU-aware RDMA communication, measured at the complete LAMMPS Loop boundary. The standard-streaming strong-scaling campaign holds the system at 1,008,000 atoms, whereas the tiled-streaming campaign uses 10,985,000 atoms; the weak-scaling campaigns use the corresponding atom count per GPU. Weak-scaling throughput speedup is defined as \(N t_1/t_N\), where \(N\) is the GPU count and \(t_N\) is Loop time per local atom.}
\label{fig:lammps_scaling_results}
\end{figure}

We test the scalability of MPI multi-GPU molecular dynamics with 1 to 64 A800 GPUs with LAMMPS.
The strong-scaling tests use a fixed total number of atoms, while the weak-scaling tests keep the number of atoms per GPU fixed.
Figure \ref{fig:lammps_scaling_results} shows that both execution modes scale nearly ideally through eight GPUs.
At 64 GPUs, the weak-scaling efficiencies are 87.92\% for standard streaming and 93.81\% for tiled streaming, corresponding to 64.51 million and 703.04 million atoms in total, respectively.
Strong-scaling efficiency decreases as the local workload contracts, reaching 45.80\% for the 1.008-million-atom standard-streaming case (15,750 atoms per GPU) and 73.91\% for the 10.985-million-atom tiled-streaming case at 64 GPUs (171,640 atoms per GPU).

\subsection{Numerical fidelity}

The numerical qualification compares Symmetrix-XL with the corresponding MACE-Torch evaluations rather than assuming equivalence from the unchanged checkpoint.
For the float32 OMAT-0 SrTiO\(_3\) task on the RTX 5090, the absolute energy-per-atom, maximum force-component, and maximum stress-component differences are \result{$1.200\times10^{-5}$~eV/atom}, \result{$6.53\times10^{-5}$~eV/\AA}, and \result{$8.82\times10^{-8}$~eV/\AA$^3$}, respectively.
Across the sampled float64 MACE-MPA-0 WBM and DynaMat configurations, energy, force, and stress differences remain within the measured tolerances, although the implementations are not bitwise identical.
Appendix~\ref{app:additional-fidelity} provides complementary curve- and trajectory-level verification, including selected diatomic curves, NVE energy conservation, radial-spline interpolation, and thermal-conductivity parity against published DFT references.

\begin{table}[htbp]
\centering
\small
\caption{Reproduction of published MACE-MPA-0 results across the WBM discovery \cite{matbench}, thermal-conductivity \cite{pota_thermal}, diatomic \cite{mace_mp0}, and DynaMat \cite{dynamat} workflows. Symmetrix-XL uses the unchanged MACE-MPA-0 medium checkpoint in float64 and follows the published task protocols.}
\label{tab:mpa0_workflow_preservation}
\begin{tabularx}{\linewidth}{@{}lYYY@{}}
\toprule
Workflow & Scope & Symmetrix-XL result & Published comparison \\
\midrule
WBM relaxation & 256,963 structures & 256,910 converged; 53 reached 500 steps & 252,373 endpoints (98.214\%) agree within \(10^{-4}\)~\AA{} \\
WBM discovery & F1; formation-energy MAE & 0.835664; 0.02760~eV/atom & 0.835569; 0.02760~eV/atom \\
Thermal conductivity & 103 calculations & SRME 0.412746; SRE 0.205094 & SRME 0.412261; SRE 0.204832 \\
Diatomic curves & 87 supported element pairs & \(5.67\times10^{-6}\)~eV/dimer MAE & Published MACE curves \\
DynaMat & 17 systems & Aggregate fidelity 0.6620 & 0.6639 \\
\bottomrule
\end{tabularx}
\end{table}

All WBM relaxations and thermal-conductivity calculations completed without runtime failure.
The small subset of WBM structures following different optimization paths changes the aggregate discovery metrics only minimally, while DynaMat agreement is assessed through published ensemble observables because independently integrated trajectories diverge.
Together, the property-level and workflow-level tests show that the optimized execution reproduces the reference model behavior relevant to the reported applications.

\section{Case studies}
\label{sec:case_studies}

We used Symmetrix-XL with pretrained MACE models to run simulations covering a range of materials, system sizes, and simulation conditions. The purpose of these examples is to test the inference system under realistic workloads, especially for large systems and molecular-dynamics runs where inference time and memory usage become limiting. We do not use these calculations to make detailed scientific claims about the individual material systems. Instead, they show the range and scale of simulations that can be carried out with Symmetrix-XL using existing pretrained models.

\subsection{Elastic response of polycrystalline alloys with varied system sizes}
\label{sec:case_lial_scale}


\paragraph{Scientific objective and simulation setup.}
The objective of this work is to quantify the finite-size effects in the calculated Young’s modulus of a nanocrystalline Al-rich Li–Al alloy and to determine the cell size required for obtaining a microstructure-representative elastic response.  The
test case is a stoichiometry-fixed nanocrystalline Li--Al alloy
(Li~16.50~at.\%, nominal density 2.39~g/cm$^3$) built as a periodic cubic
Voronoi polycrystal (random-seed tessellation via
SciPy~\cite{scipy2020}) with 24 grains, consisting of 8 grains drawn from a LiAl$_3$-type
source structure and 16 from a LiAl$_7$-type source structure, with target
phase volume fractions 31.84\% and 68.16\%.

To isolate the finite-size effect, the same microstructure generation
protocol was applied to four cubic simulation cells with edge lengths of
69.5, 118.5, 255.0, and 548.0~\AA{}, corresponding to approximately
$2.0\times10^4$, $1.0\times10^5$, $1.0\times10^6$, and $1.0\times10^7$
atoms. All atomic interactions were described using the Symmetrix-XL~\cite{symmetrixxl}
v0.1.1 MACE machine-learning potential~\cite{mace2022} for the Li--Al system. After
structural relaxation, uniaxial tensile simulations were performed at
300~K along three orthogonal directions under periodic boundary
conditions. The axial strain was increased from 0 to 4\%, while lateral
dimensions were adjusted to minimize transverse stresses. Young's modulus
was extracted from the linear stress--strain response, with the 0--2\%
strain range used as the primary elastic fitting window and the 0--4\%
range used for sensitivity analysis.

\begin{figure}[!htbp]
\centering
\includegraphics[width=\textwidth]{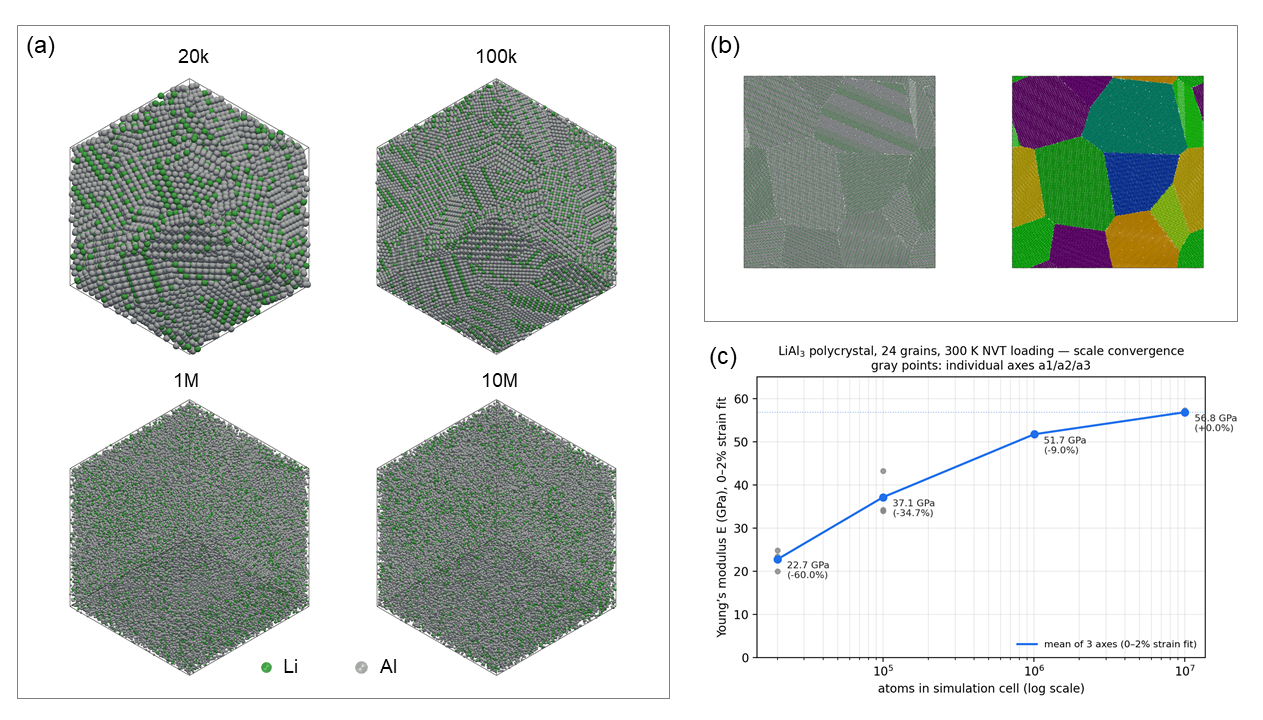}
\caption{\textbf{(a)} The four sample cells of the scale series (same 24-grain
construction, Li green, Al silver). Left to right / top to bottom: 20k, 100k,
1M, and 10M atoms. The 1M and 10M panels show a uniformly subsampled subset
of atoms (1/5 and 1/50, display radii rescaled for legibility); the 20k and
100k panels show every atom. The 10M cell edge (548~\AA) is $7.9\times$ the
20k edge at identical composition and grain count. \textbf{(b)} True-scale 4~\AA{}-thick slice through the 10M cell
($z \in [270, 274]$~\AA{}, 73{,}005 atoms). Left: element coloring showing
differently oriented grains and disordered boundary atoms. Right: the same
slice colored by grain ID, showing the Voronoi tessellation (LiAl$_3$-type
grains $\times$8, LiAl$_7$-type $\times$16) with jagged grain-boundary
facets. \textbf{(c)} Direction-averaged 0--2\% Young's modulus versus atom count (blue
line and points; gray points are the three individual axes $a_1$--$a_3$,
from the archived per-axis fits). Annotations give each mean and its
deviation from the 10M value. The 0--2\% series has not
saturated at $10^7$ atoms.}
\label{fig:lial_structures}
\label{fig:lial_slab}
\label{fig:lial_convergence}
\end{figure}

\paragraph{Observed response.}
For the 10M cell the axial stress rises from a small residual compressive
value at $\varepsilon=0$ (intercept $-0.32$ to $-0.41$~GPa, the fixed-length
thermal reference state) to $0.71$--$0.74$~GPa at 2\% strain and
$1.54$--$1.58$~GPa at 4\% strain (Table~\ref{tab:lial_scale}). The
0--2\% fits give $E_{a_1}=56.53$, $E_{a_2}=56.62$, $E_{a_3}=57.30$~GPa
($R^2 \ge 0.9989$), a direction-averaged modulus of
$E = 56.8 \pm 0.4$~GPa (half-range across axes) with an anisotropy spread of
only 1.3\%, as expected for a randomly oriented polycrystal. Extending the
fit window to 0--4\% lowers the apparent modulus to $49.3 \pm 0.5$~GPa
($R^2 \approx 0.992$), i.e.\ the stress--strain curve is measurably
sublinear beyond $\sim$2\% strain; consistent with this, the measured
effective Poisson ratio drifts from 0.300 to 0.314 between 0 and 4\% strain.
The axial stress standard error of the sampled mean is
$2$--$6\times10^{-4}$~GPa and the sampling temperature is
$300.09 \pm 0.03$~K in every stage, so the scatter between axes reflects
physical fluctuation of the finite cell rather than thermal drift. Across the
scale series (Fig.~\ref{fig:lial_convergence},
Table~\ref{tab:lial_scale}), the direction-averaged 0--2\% modulus increases
monotonically from 22.7~GPa (20k) to 37.1~GPa (100k) to 51.7~GPa (1M) to
56.8~GPa (10M), while the axis-to-axis spread collapses from $\sim$21\% at
20k to 1.3\% at 10M and the fit quality improves from $R^2 = 0.78$--$0.86$
to $R^2 \ge 0.999$ (Fig.~\ref{fig:lial_structures} shows the four sample
cells; Fig.~\ref{fig:lial_slab} resolves the 10M grain-boundary network at
true atomic scale). The monotonic rise and the shrinking anisotropy are
observations; attributing the rise to the dilution of grain-boundary
softening~\cite{schiotz1998,meyers2006} and the sublinearity beyond 2\% to
grain-boundary--mediated inelasticity~\cite{vanswygenhoven2004} is
interpretation imported from the nanocrystalline-plasticity literature, not a
mechanism demonstrated within these trajectories (see the last paragraph).

\begin{table}[t]
\centering
\small
\begin{tabular}{rrrrrrrr}
\toprule
Atoms & Edge (\AA) & \multicolumn{3}{c}{$E$, fit 0--2\% (GPa)} &
\multicolumn{2}{c}{$\bar{E}$ (GPa)} & \\
\cmidrule(lr){3-5}\cmidrule(lr){6-7}
$N$ & $L$ & $a_1$ & $a_2$ & $a_3$ & 0--2\% & 0--4\% & grain $d$ (nm) \\
\midrule
19{,}927      & 69.5  & 24.78 & 23.39 & 20.00 & 22.72 & 19.72 & 2.4  \\
99{,}886      & 118.5 & 43.23 & 34.19 & 33.93 & 37.12 & 28.34 & 4.1  \\
1{,}004{,}260 & 255.0 & 52.04 & 51.44 & 51.66 & 51.71 & 39.64 & 8.8  \\
10{,}009{,}503& 548.0 & 56.53 & 56.62 & 57.30 & 56.82 & 49.33 & 19.0 \\
\bottomrule
\end{tabular}
\caption{Scale series for the 24-grain Li--Al polycrystal at 300~K.
$E$ is the least-squares slope of the axial stress versus applied strain in
the stated window; $\bar{E}$ is the mean over the three axes. The nominal
grain diameter $d$ scales as $L/24^{1/3}$. Data from
\texttt{modulus\_300K\_summary.json} of each case; the 10M-row slopes were
independently recomputed from the raw per-stage stress--strain CSVs.}
\label{tab:lial_scale}
\end{table}

\paragraph{Role of system scale.}
The scale study isolates one specific finite-size effect: with the grain count
fixed at 24, the grain diameter scales with the cell edge,
$d \propto L/24^{1/3}$, from 2.4~nm (20k) to 19~nm (10M), and the
grain-boundary area per unit volume scales as $1/d$. The 10M cell is the
first in the series in which this dilution is quantitatively separated from
statistical noise, for three reasons. First, each grain contains
$8 \times 10^2$ to $4.2 \times 10^5$ atoms across the series; only in the
two largest cells do individual grains sample their local stress average with
standard errors ($\le 6\times10^{-4}$~GPa) far below the modulus differences
being resolved. Second, the elastic isotropy of a 24-grain random polycrystal
is itself a finite-size observable: the axis-to-axis spread of $E$ falls from
21\% (20k) to 1.3\% (10M), so the 10M cell is the smallest one in which the
three loading directions give a direction-invariant modulus rather than three
noise realizations. Third, the cell edge of the 10M model (54.8~nm) exceeds
the grain diameter by a factor of $2.9$, so the cell contains the complete
grain-boundary network --- dozens of facets up to $\sim$50~nm long with
triple lines and quadruple junctions --- embedded in grain interiors that are
themselves resolved at atomic detail
(Fig.~\ref{fig:lial_slab}). The scientific payoff is visible in the numbers:
the modulus increment from 1M to 10M atoms is still $+5.1$~GPa ($+9.9$\%),
so a 1M-atom cell --- already large by polycrystal-MD standards --- would
have incorrectly treated the residual finite-size softening as converged and
understated the modulus by 9\%. At 10M the same construction can also carry
the full loading protocol (three replicas, five strain stages, 83.4~ps of NVT
dynamics) in $\sim$103~h, which is what makes the scale-resolved
ladder affordable rather than anecdotal.

\paragraph{Validation and limitations.}
The numerical consistency was assessed through fitting-window comparison,
directional reproducibility, stress sampling, and structural relaxation.
The 0--2\% fit gives a modulus approximately 15\% higher than the 0--4\%
fit, while the three loading directions of the 10M-atom cell differ by only
1.3\%. The stress sampling uncertainty is below $6\times10^{-4}$~GPa, the
temperature remains at $300.10\pm0.03$~K, and the final RMS force after
relaxation is approximately $5.6\times10^{-4}$~eV/\AA{}.

These checks establish the internal consistency of the simulations but do
not quantify model uncertainty. The calculations include only one Voronoi
realization per scale, use geometrically generated grain boundaries without
annealing, and consider only tensile loading along three axes. Potential
uncertainty, precision effects, and microstructure variability are not
included.

\FloatBarrier


\subsection{Lithium transport and vibrational dynamics at a cathode/electrolyte interface}
\label{sec:li-interface}

\begin{figure}
    \centering
    \includegraphics[width=\linewidth]{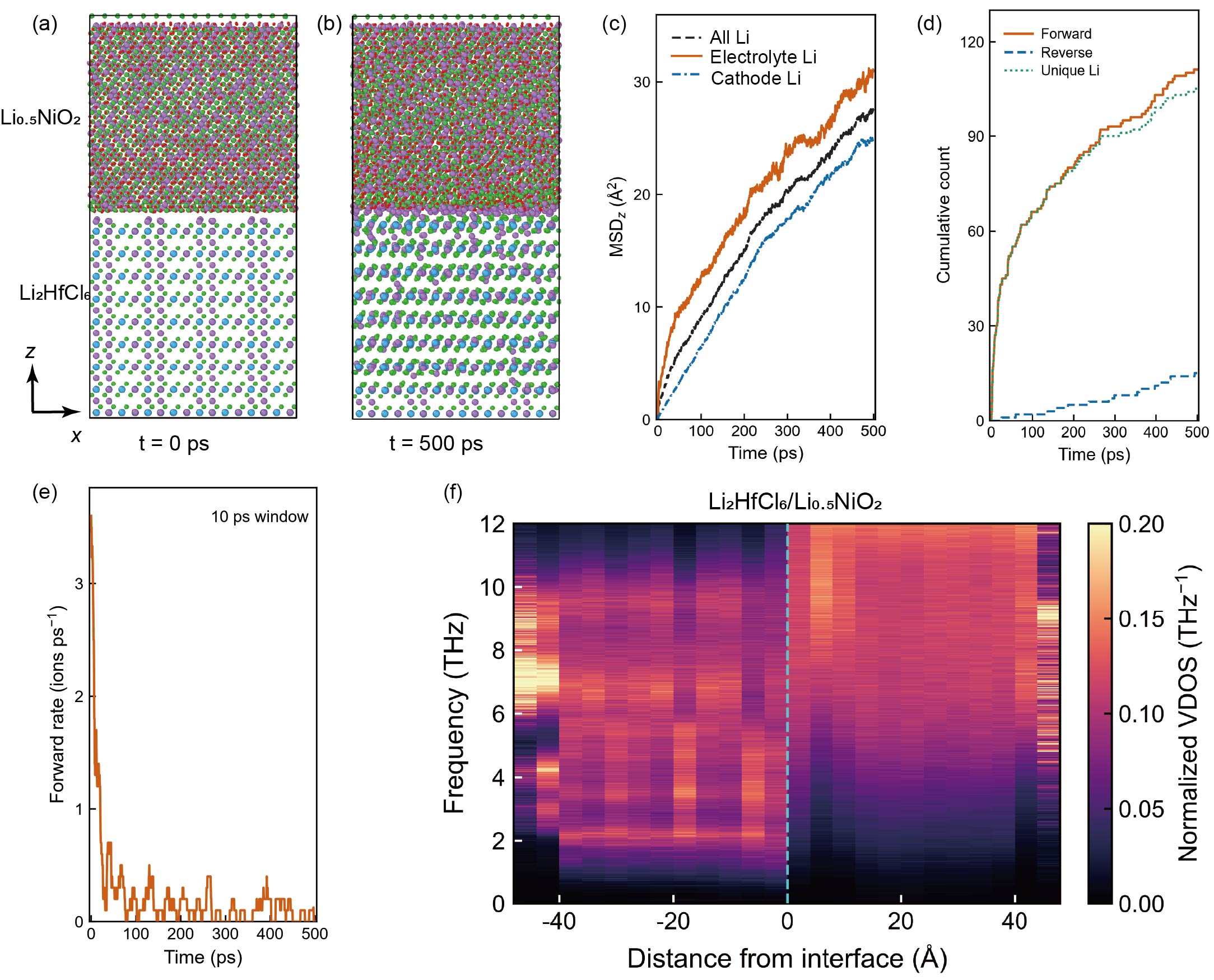}
    \caption{Lithium transfer and spatially resolved vibrational spectra at the $\chem{Li_{0.5}NiO_2}(104)\,|\,\chem{Li_2HfCl_6}(100)$ interface.
    (a,b) Atomic configurations at $t=\SI{0}{\pico\second}$ and $t=\SI{500}{\pico\second}$ during the NVT simulation at \SI{600}{K}, with the cathode above the electrolyte.
    (c) Drift-corrected MSD normal to the interface for all mobile Li ions and for the subsets initially located in the electrolyte and cathode.
    (d) Cumulative counts of electrolyte-to-cathode transfers (Forward), cathode-to-electrolyte transfers (Reverse), and distinct electrolyte-origin Li ions that have entered the cathode at least once (Unique Li).
    (e) Forward-transfer rate evaluated in a \SI{10}{\pico\second} moving window.
    (f) Spatially resolved all-atom VDOS from the \SI{100}{\pico\second} NVE trajectory initialized at the NVT endpoint.
    Spectra are calculated in \SI{4}{\angstrom}-wide bins and normalized to unit integrated area over \SIrange{0}{12}{\tera\hertz}.
    Negative and positive distances correspond to the electrolyte and cathode sides, respectively; the cyan dashed line marks the interface.
    The color scale spans \SIrange{0}{0.20}{\per\tera\hertz}, with larger values displayed at the upper color limit.}
    \label{fig:li-interface}
\end{figure}

\paragraph{Scientific objective and simulation setup.}
We examined Li transfer and local vibrational dynamics at the $\chem{Li_{0.5}NiO_2}(104)\,|\,\chem{Li_2HfCl_6}(100)$ interface in a \num{11728}-atom model.
The simulations used the fine-tuned MACE potential, which was initialized from the medium MACE-MP-0 foundation checkpoint~\cite{mace2022,mace_mp0}, evaluated in FP32 precision with the Symmetrix inference engine~\cite{kovacs2025,mace_mp0,symmetrix_original}.
The pre-relaxed orthogonal cell measured approximately $32.88 \times 50.18 \times 99.30$~\AA$^{3}$ and had an interfacial area of \SI{1649.91}{\angstrom\squared}.
We ran a \SI{500}{\pico\second} NVT simulation at \SI{600}{K} with the Bussi thermostat~\cite{bussi2007}, a \SI{1}{\femto\second} time step, and a thermostat coupling time of \SI{100}{\femto\second}.
Atoms within \SI{5.8}{\angstrom} of the two boundaries normal to the interface were held fixed, leaving \num{10508} mobile atoms.
Panels (a) and (b) of Figure~\ref{fig:li-interface} show the initial and final configurations.
The final positions and velocities were then used to initialize a \SI{100}{\pico\second} NVE trajectory for spatially resolved vibrational density of states (VDOS) analysis.

\paragraph{Observed response.}
In the present trajectory, Li transfer occurred predominantly from the electrolyte into the cathode.
Over \SI{500}{\pico\second}, \num{111} forward and \num{15} reverse events gave \num{96} net forward events (Figure~\ref{fig:li-interface}d), equivalent to a net areal flux of $11.64~\mathrm{ions\,nm^{-2}\,ns^{-1}}$.
Of the \num{105} distinct electrolyte-origin Li ions that entered the cathode side at least once, \num{64} remained cathode-like at the end of the trajectory.
At \SI{500}{\pico\second}, the drift-corrected mean-square displacement (MSD) normal to the interface was \SI{30.98}{\angstrom\squared} for electrolyte-origin mobile Li and \SI{24.78}{\angstrom\squared} for cathode-origin mobile Li (Figure~\ref{fig:li-interface}c).
Although the final MSD values differed, linear fits over \SIrange{100}{500}{\pico\second} gave similar one-dimensional tracer diffusion coefficients: \SI{2.21e-6}{\centi\metre\squared\per\second} and \SI{2.29e-6}{\centi\metre\squared\per\second}, respectively.
Within this trajectory, the forward-event rate, evaluated in a \SI{10}{\pico\second} moving window, exhibits rapid initial transfer followed by slower, intermittent exchange (Figure~\ref{fig:li-interface}e).
In the VDOS map, distinct bands appear at approximately \SIrange{2}{7}{\tera\hertz} on the electrolyte side (Figure~\ref{fig:li-interface}f).
The cathode side has a broader distribution of spectral weight extending towards \SI{12}{\tera\hertz} and weaker intensity at low frequencies.

\paragraph{Role of system scale.}
At this system size, the analysis includes \num{105} distinct electrolyte-origin Li ions that cross the interface, with separate counts for ions and crossing events.
The cell spans both materials, allowing the local VDOS to be resolved in \SI{4}{\angstrom}-wide bins across the interface.
The \SI{500}{\pico\second} trajectory captures the change from rapid initial transfer to slower, intermittent exchange.

\paragraph{Validation and limitations.}
The main limitation is that the model assumes an ideal, defect-free interface and does not explicitly account for a space-charge layer.
The results therefore describe Li exchange under these idealized conditions; the effects of interfacial defects and space-charge formation are outside the scope of this study.
The transport analysis accounts for framework drift and rapid recrossings, while the normalized VDOS maps compare relative spectral distributions.
The reported transport statistics are obtained from a single \SI{500}{\pico\second} NVT trajectory and therefore represent one stochastic realization of the interfacial dynamics. Accordingly, the observed preference for electrolyte-to-cathode Li transfer and its temporal evolution are interpreted as trajectory-level behavior under the present simulation conditions, rather than as a statistically converged kinetic rate or a universal mechanistic characteristic of the interface.

\FloatBarrier

\subsection{Shock-driven decomposition of a \texorpdfstring{$\beta$}{beta}-HMX crystal}
\label{sec:hmx-case}

\paragraph{Scientific objective and simulation setup.}
We shock-loaded a \num{201600}-atom $\beta$-HMX crystal (\num{7200} \chem{C_4H_8N_8O_8} molecules; $98.5 \times 173.4 \times 108.0$~\AA$^{3}$ at $t = 0$) with a sinusoidal travelling-wave force field of amplitude \SI{0.15}{eV\per\angstrom}, phase velocity \SI{20}{\angstrom\per\ps}, and wavelength \SI{5}{\angstrom}.
The crystal was initialized at \SI{600}{K} and integrated in LAMMPS~\cite{lammps2022} through the Symmetrix-XL pair style with the MACE-OFF23 potential~\cite{kovacs2025}, in the shock-driven NPH ensemble ($P_0 = 0$) so that the reacting material expands freely.

\begin{figure}[!htbp]
\centering
\includegraphics[width=\textwidth]{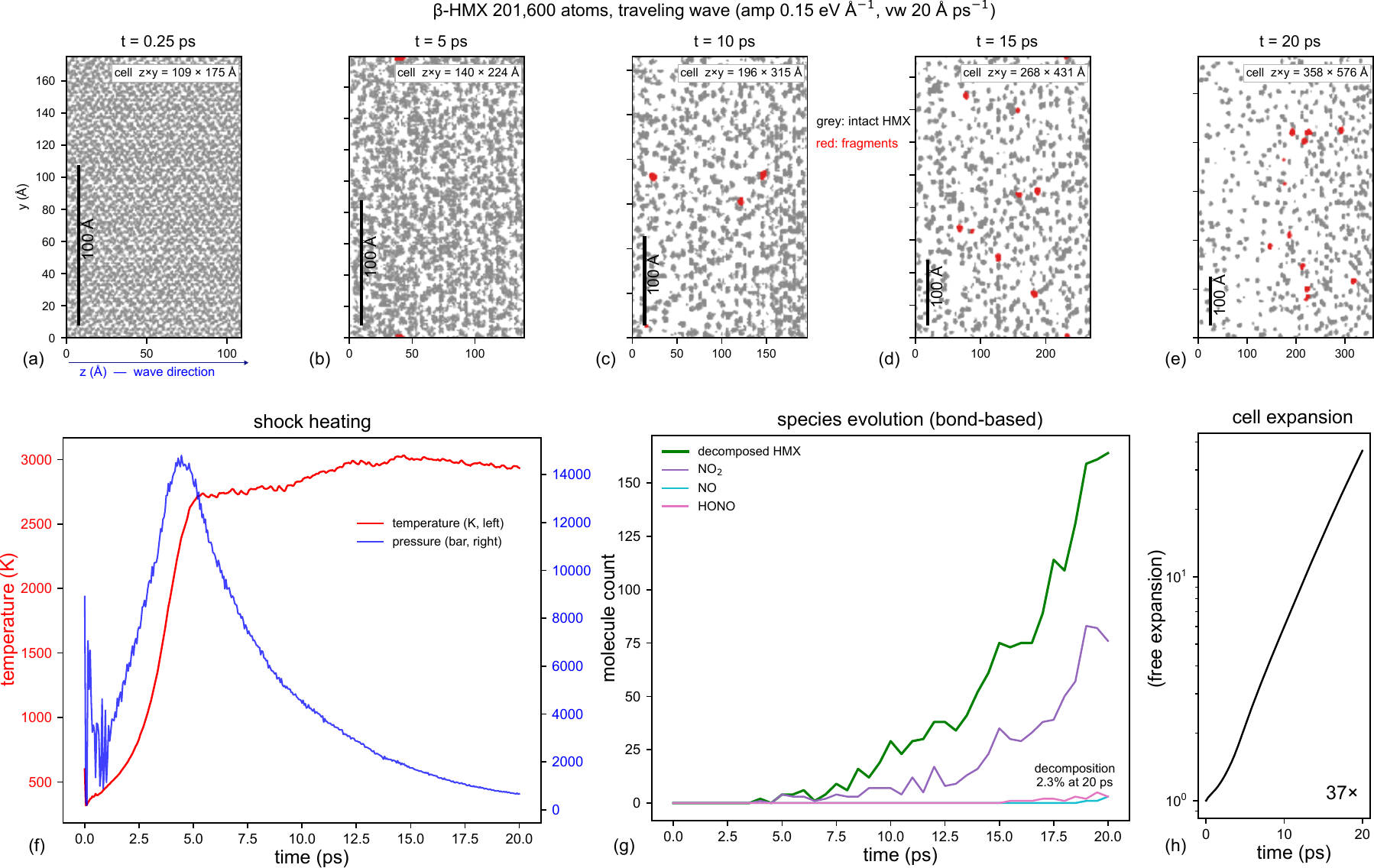}
\caption{Shock-driven decomposition of a \num{201600}-atom $\beta$-HMX crystal (MACE-OFF23; $98.5 \times 173.4 \times 108.0$~\AA$^{3}$ at $t = 0$) driven by a travelling wave (\SI{0.15}{eV\per\angstrom}, \SI{20}{\angstrom\per\ps}, $\lambda = \SI{5}{\angstrom}$) at \SI{600}{K} under NPH conditions ($P_0 = 0$).
\textbf{(a--e)} Chemical-damage maps at $t = 0.25$, 5, 10, 15, and \SI{20}{ps}; each panel shows a slab of the cell (fractional $x < 0.06$) spanning the full instantaneous $y$--$z$ cross-section, with the propagation direction $z$ horizontal.
Grey denotes intact HMX molecules; red denotes decomposition fragments and products classified from the bond graph.
Each panel is drawn at its own instantaneous cell size, labelled with the current cell dimensions, and carries a fixed 100~\AA{} scale bar.
\textbf{(f)} Temperature (red, left axis) and pressure (blue, right axis).
\textbf{(g)} Species evolution from the bond-graph analysis: decomposed HMX molecules and the products \NOtwo{}, NO, and \HONO{} on a single molecule-count axis.
\textbf{(h)} Relative cell volume on a logarithmic scale.}
\label{fig:hmx}
\end{figure}

\paragraph{Observed response.}
The drive converts the initially crystalline explosive into a hot, expanding, partially dissociated fluid (Figure~\ref{fig:hmx}a--e).
The temperature rises to $\approx\SI{3000}{K}$ within \SI{5}{ps} and then changes only slowly (Figure~\ref{fig:hmx}f), while the cell dilates $\approx37$-fold (Figure~\ref{fig:hmx}h).
Decomposition proceeds by nitro-group elimination and hydrogen loss, with \NOtwo{} and H as the earliest products (Figure~\ref{fig:hmx}g); 
this product sequence constitutes contextual qualitative evidence consistent with the ReaxFF studies of nitramines~\cite{strachan2005} (but does not serve as quantitative validation of MACE-OFF23 for shocked \(\beta\)-HMX).
By \SI{20}{ps}, 164 of the \num{7200} molecules ($2.3\%$) have decomposed, and the rate accelerates markedly after $\approx\SI{10}{ps}$ as the temperature and the concentration of reactive fragments increase.

\paragraph{Role of system scale.}
Two features of the chemistry require the \num{7200}-molecule cell.
The first is that scission is collective: a molecule breaks within the pressure and temperature field of its neighbours, so both the product hierarchy and the onset of acceleration are ensemble properties rather than single-molecule outcomes.
The second is statistical: those ensemble properties must be sampled over a meaningful number of events, and the 164 decomposition events collected here are what allow the product hierarchy and the onset of acceleration to be quantified.

\paragraph{Validation and limitations.}
The shock-loading protocol, ensemble definitions, bond-graph reaction-identification thresholds, and product-counting rules are fixed and recorded in the input and analysis scripts used to generate this trajectory.
Two checks bound the interpretation of these results.
The first concerns the time window: mapping the drive strength onto the accessible window---\SI{0.85}{ps} at \SI{0.5}{eV\per\angstrom} versus more than \SI{20}{ps} at \SI{0.15}{eV\per\angstrom}---shows that the usable simulation time is set by the drive and by the thermodynamic state reached by the potential, not by the computational budget alone.
The second is numerical: the endpoint at $\approx\SI{20.4}{ps}$ is reproduced under fixed, halved, and adaptive (\texttt{dt/reset}) time steps, indicating that it is not a time-step artefact.
With only $2.3\%$ of the molecules decomposed by the endpoint, the trajectory resolves the onset of decomposition rather than the later stages of product chemistry, and the results are specific to the loading geometry and drive used here.

The $\beta$-HMX case shows that a machine-learned potential can carry a \num{201600}-atom shock simulation through the onset of decomposition and resolve a product sequence consistent with the cited ReaxFF study, with the accessible window set by the drive and the thermodynamic state rather than by computational cost.

\FloatBarrier

\subsection{Ring-opening polymerization and S--S crosslinking in a thiirane (\texorpdfstring{\chem{C_2H_4S}}{C2H4S}) liquid}
\label{sec:c2h4s-case}

\paragraph{Scientific objective and simulation setup.}
We simulated an approximately 200k-atom thiirane liquid containing \num{28571} \chem{C_2H_4S} molecules, 10 \chem{CH_3S} spectators, and 10 TMA spectators with the MACE-OFF23 potential~\cite{kovacs2025}. Simulations were performed using a standard LAMMPS backend, MACE pair style, double numerical precision, and an NPT ensemble; the system was held at \SI{1500}{K} for \SI{16.75}{ps} at nearly constant volume.
The strong-damping barostat self-pressurises the cell to $\approx\SI{12}{GPa}$ as the initially expanded liquid contracts.

\begin{figure}[!htbp]
\centering
\includegraphics[width=\textwidth]{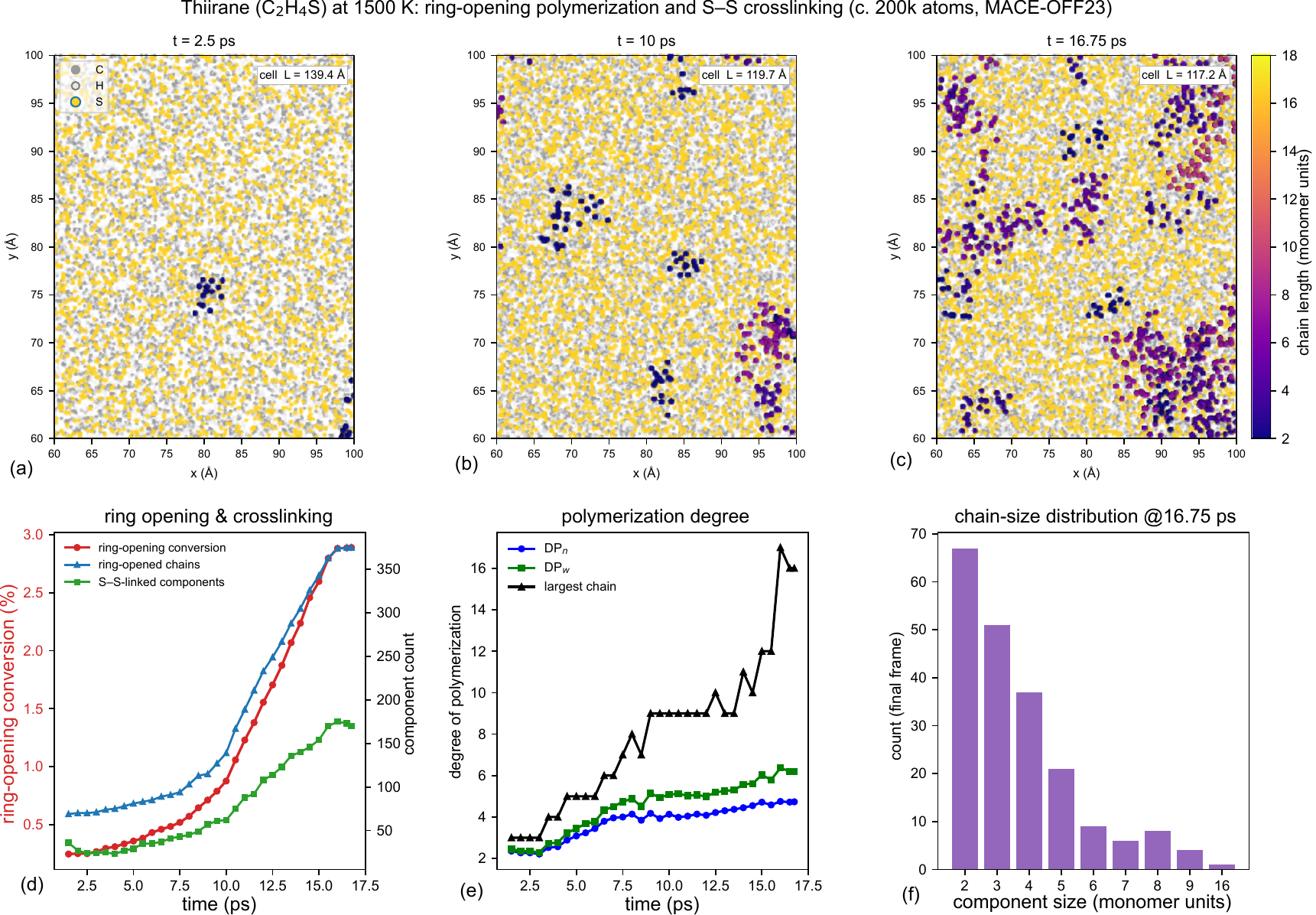}
\caption{Ring-opening polymerization and S--S crosslinking in an approximately 200k-atom thiirane liquid containing \num{28571} \chem{C_2H_4S} molecules, 10 \chem{CH_3S} spectators, and 10 TMA spectators (MACE-OFF23, \chem{C/H/S}) at \SI{1500}{K} over \SI{16.75}{ps}; connectivity and ring counts come from the bond graph ($E - V + 1$ cycles per component).
\textbf{(a--c)} Snapshots at $t = 2.5$, 10, and \SI{16.75}{ps}: a fixed $40 \times 40$~\AA{} sub-window in absolute cell coordinates, with a 20~\AA{} scale bar and the instantaneous cubic cell edge $L$ ($139.4$, $119.7$, and \SI{117.2}{\angstrom}).
Grey denotes carbon, white hydrogen, and yellow sulfur; atoms in chains of two or more units are coloured by the number of units per component (colour bar, 2--18).
\textbf{(d)} Ring-opening conversion (red, left axis), ring-opened chain species (blue), and S--S-linked components (green, right axis).
\textbf{(e)} Degree of polymerization: $\mathrm{DP}_n$ (blue), $\mathrm{DP}_w$ (green), and largest chain (black).
\textbf{(f)} Chain-size distribution at \SI{16.75}{ps}.
Panels (d)--(f) use a uniform \SI{0.5}{ps} grid so that neither trajectory segment is over-weighted.}
\label{fig:c2h4s}
\end{figure}

\paragraph{Observed response.}
Ring-opening conversion is negligible at first ($0.25\%$ at \SI{1.5}{ps}), accelerates steadily to $2.89\%$, and then plateaus (Figure~\ref{fig:c2h4s}d).
Chain growth accompanies the conversion: the number- and weight-average degrees of polymerization rise from 2.3 to 4.7 and from 2.4 to 6.3, respectively, and the largest chain grows from 3 to 16 units (Figure~\ref{fig:c2h4s}e).
The final chain-size distribution is dominated by dimers and trimers with a tail of up to 16 units (Figure~\ref{fig:c2h4s}f).

\paragraph{Role of system scale.}
All chemical and statistical results reported herein are obtained from a single independent trajectory, with no additional replica simulations performed. Both mechanistic channels expected from thiirane chemistry~\cite{sigwalt1976,quirk2007} are resolved in a single trajectory: \CS{} ring opening ($\approx375$ chain species at the plateau) and intermolecular S--S coupling ($\approx180$--$345$ species, depending on stage), the latter counted topologically as the number of components carrying a sulfur--sulfur linkage.
The two channels are also coupled in time, since chain growth near \SI{15.8}{ps} proceeds by merging existing chains rather than by initiation: the number of distinct ring-opened chains saturates at $\approx375$ while the largest component grows from 10 to 16 units.
Only with \num{28571} monomers do the conversion, the branching ratio between the two channels, and the chain-length distribution become population quantities, each sampled across hundreds of events.

\begin{minipage}{\textwidth}
\paragraph{Validation and limitations: bond-criterion robustness.}
\label{sec:sscriterion}
Of the quantities analysed here, only the S--S topology depends on an arbitrary bond definition, so we tested its sensitivity explicitly.
The S--S distance distribution is bimodal, with a covalent peak at $\approx 2.0$--\SI{2.1}{\angstrom} and a soft non-bonded cloud peaking at $2.6$--\SI{2.7}{\angstrom}, the latter coinciding with the conventional $1.25 \times (r_{\mathrm{S}} + r_{\mathrm{S}}) = \SI{2.625}{\angstrom}$ criterion.
Applied naively, that criterion counts hundreds of weak contacts as bonds and produces an apparent collapse of the S--S-linked population from $\approx350$ to $\approx180$ components between \SI{15.5}{} and \SI{16.0}{ps}---precisely the window in which the largest chain grows to 16 units.
Stricter criteria show the collapse to be a threshold artefact (Table~\ref{tab:ss}): the tight population is constant at $\approx200$ bonds, whereas the count at \SI{2.625}{\angstrom} fluctuates by $\approx40\%$.
All results above therefore use a fixed \SI{2.45}{\angstrom} criterion, under which the S--S component count increases monotonically and the conversion and chain-length statistics are unchanged ($2.89\%$, $\mathrm{DP}_n = 4.7$, and 16 units at \SI{16.75}{ps}).
The conventional $1.25 \times (r_i + r_j)$ criterion is retained for \CH{} and \CS{} bonds, which show no comparable ambiguity.
Ring counts are read from the bond graph as $E - V + 1$ cycles per connected component, so that an opening event requires an actual ring rupture, and panels (d)--(f) are evaluated on a uniform \SI{0.5}{ps} grid so that neither trajectory segment is over-weighted.
Two caveats bound the interpretation of these statistics: the analysed endpoint requires the halved time step, since at the full step the trajectory reaches only $\approx\SI{15.1}{ps}$, and at $2.89\%$ conversion the system remains far from complete polymerization, so the distributions describe early-stage network formation rather than a converged product.

\begin{center}
\captionof{table}{Number of S--S bonds in the thiirane trajectory under different covalent criteria.
The tight population is stable, whereas the conventional $1.25 \times$ radii criterion fluctuates by $\approx40\%$; a fixed \SI{2.45}{\angstrom} criterion is therefore used throughout.}
\label{tab:ss}
\begin{tabular}{l S[table-format=4.0] S[table-format=4.0] S[table-format=4.0]}
\toprule
{S--S criterion} & {\SI{15.50}{ps}} & {\SI{15.875}{ps}} & {\SI{16.75}{ps}} \\
\midrule
$< \SI{2.30}{\angstrom}$ & 197 & 206 & 202 \\
$< \SI{2.50}{\angstrom}$ & 209 & 210 & 205 \\
$< \SI{2.625}{\angstrom}$ ($1.25 \times$ radii) & 359 & 260 & 218 \\
$< \SI{2.80}{\angstrom}$ & 1804 & 1464 & 677 \\
\bottomrule
\end{tabular}
\end{center}
\end{minipage}

The thiirane case shows that one trajectory containing \num{28571} thiirane molecules resolves both expected reaction channels of thiirane---ring-opening polymerization and S--S crosslinking---and yields conversion, branching, and chain-length statistics that are robust once the S--S topology is defined by a covalent criterion rather than by the conventional contact cutoff.

\FloatBarrier

\subsection{SEI Full-Cell Atomic-Scale Simulation}
\label{sec:sei-full-cell}

\begin{figure}[H]
\centering
\includegraphics[width=\textwidth,height=0.92\textheight,keepaspectratio]{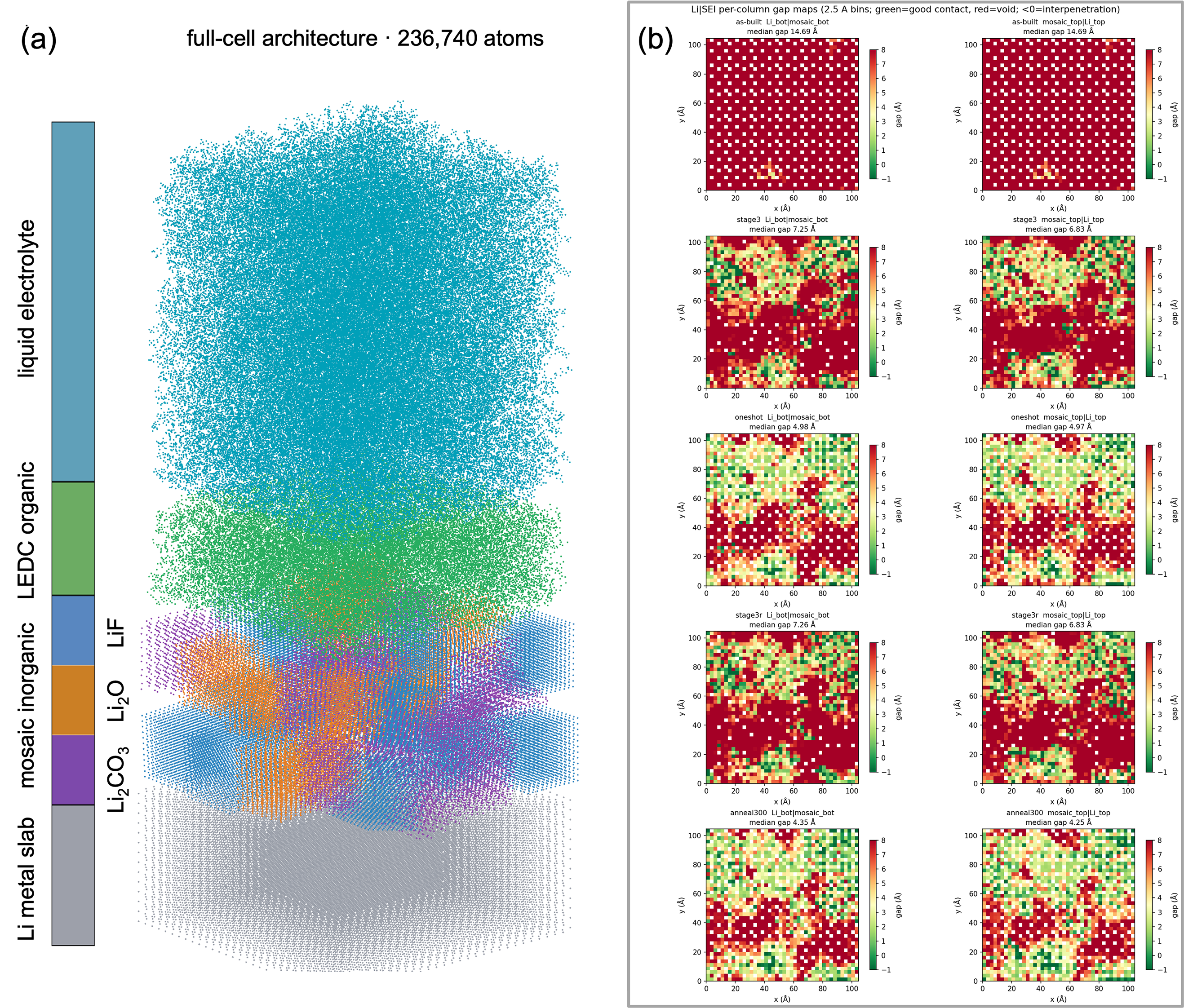}
 \caption{\textbf{(a)} Full-cell architecture (schematic render, lower four-phase stack: Li metal slab | mosaic inorganic layer | \chem{LEDC} organic layer | liquid electrolyte; the top mirror half is omitted for clarity). Grey = Li metal; dark blue = \chem{LiF} grains; dark orange = Li$_{2}$O grains; purple = Li$_{2}$CO$_{3}$ grains (43 grains per mosaic layer); green = \chem{LEDC} organic SEI; teal = liquid electrolyte (EC:DMC + 1.2 M LiPF$_{6}$). \textbf{(b)} Column-resolved contact analysis of the relaxed Li | mosaic interfaces. Top: gap evolution; middle: interfacial column contact maps (green = contact column, red = void column) --- the island-like pattern is visible to the eye; bottom: nearest-neighbour distance spectrum.}
\label{fig:f1}
 \end{figure}

\paragraph{Scientific objective and simulation setup.}
We investigated the relationship between chemical heterogeneity and Li transport in a model solid--electrolyte interphase (SEI) using a 358,052-atom multilayer cell comprising Li metal, a 43-grain mosaic inorganic layer (\chem{LiF}/Li$_{2}$O/Li$_{2}$CO$_{3}$), lithium ethylene dicarbonate (\chem{LEDC}), and liquid EC/DMC electrolyte containing LiPF$_{6}$ (Figure~\ref{fig:f1}). The mosaic was assembled by best-effort Packmol packing~\cite{packmol2009} and healed by MLIP-controlled relaxation. The simulations used a fine-tuned MACE-OMAT foundation potential~\cite{mace2022,mace_mp0} with Li--P--H--O--C--F coverage --- a machine-learning interatomic potential evaluated with the Symmetrix inference engine (FP32 on the full cell; FP64 on the transport line). The pre-relaxed orthogonal cell measured 105.27 $\times$ 105.27 $\times$ 455.7~\AA{}$^{3}$ (cross-sectional area 11,081.8~\AA{}$^{2}$). Relaxation used staged positional release under Langevin dynamics, followed by a comparison of staged relaxation, elevated-temperature annealing, and all-released relaxation. A bond-aware structural guard was used to avoid classifying covalent intramolecular distances as nonbonded overlaps. Transport analysis used four NVT trajectories with temperatures of 300.0, 303.2, 376.3, and 449.2 K, with production segments of 6.1--8.0 ps.

\begin{figure}[htbp]
 \centering
 \includegraphics[width=\textwidth,height=0.92\textheight,keepaspectratio]{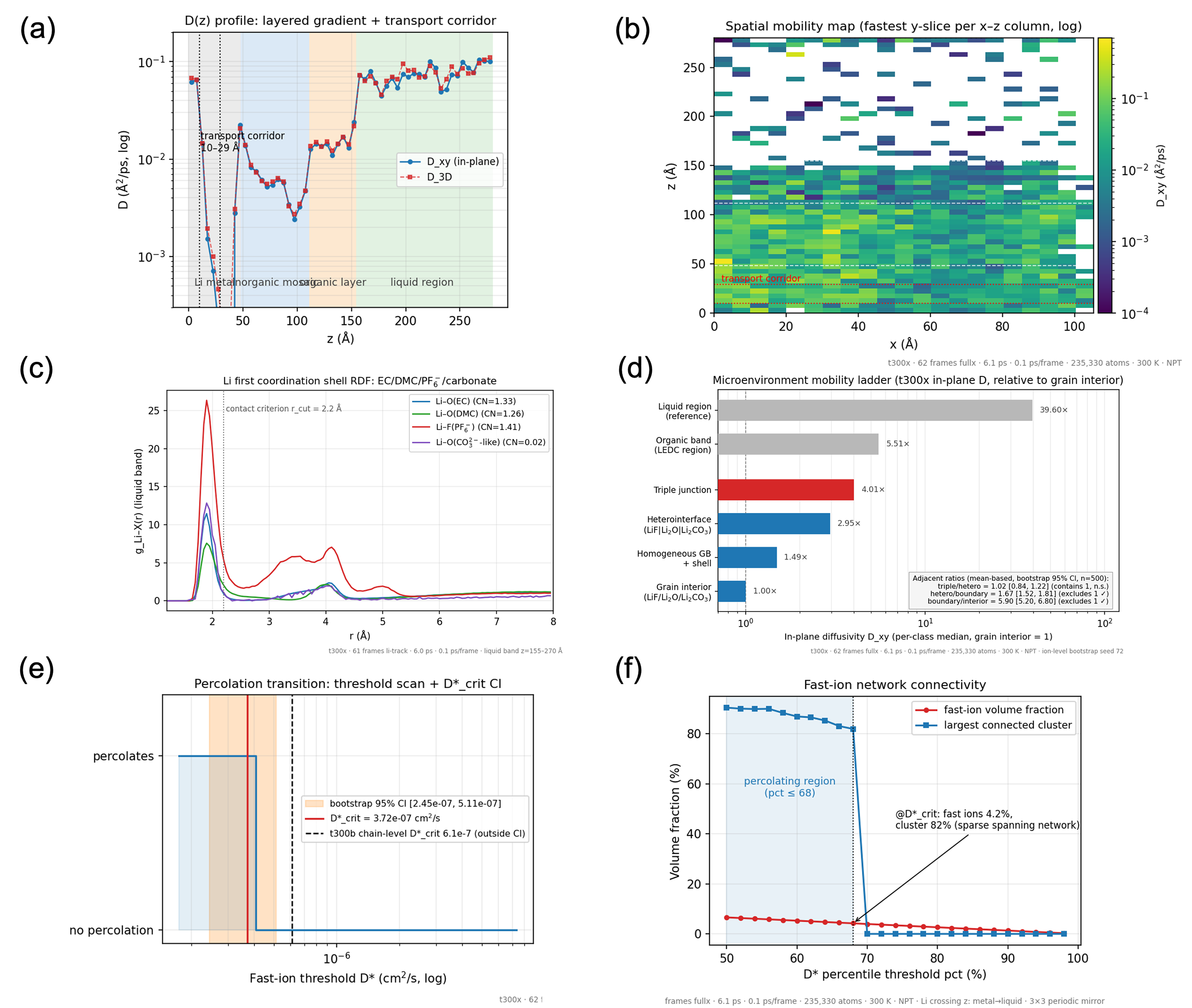}
\caption{\textbf{(a)} $D_{\mathrm{Li}}$(z) profile (in-plane $D_{\mathrm{xy}}$ and 3D $D_{\mathrm{3D}}$ conventions) with four-zone shading; the dotted line marks the fast-transport corridor (z = 50--145~\AA{}). \textbf{(b)} spatial mobility map. \textbf{(c)} Liquid-band Li first-shell RDF. \textbf{(d)} Microenvironment mobility ladder. \textbf{(e)} Spanning-probability threshold scan (step function) with the $D^{*}_{\mathrm{crit}}$ confidence band (orange, 3.72(+1.39/-1.27)$\times$10$^{-7}$ cm$^{2}$/s) and the chain-level point estimate (black dashed). \textbf{(f)} Fast-ion volume fraction and largest-connected-cluster fraction vs. threshold percentile.}
\label{fig:f2}
\end{figure}

\paragraph{Observed response.}
The in-plane $D_{\mathrm{Li}}(z)$ profile varied substantially across the layered model, with values of 4.8--5.2$\times$10$^{-6}$ cm$^{2}$/s in the liquid band and 1.3--2.0$\times$10$^{-6}$ cm$^{2}$/s in the \chem{LEDC} transition band (Figure~\ref{fig:f2}(a,b)). The lower values reported across the inorganic mosaic should be interpreted as apparent atomic-mobility estimates under the chosen trajectory and spatial binning; Li-atom self-diffusion in the metallic slab is not equivalent to Li$^{+}$ transport through the SEI. Within the mosaic, homophase grain boundaries were 5.90 times as mobile as grain interiors (bootstrap 95\% CI: 5.20--6.80), while heterophase interfaces were a further 1.67 times as mobile (1.52--1.81). The triple-junction-to-heterophase-interface ratio was 1.02 (0.84--1.22), providing no statistically resolved evidence of additional enhancement (Figure~\ref{fig:f2}(d)).

Voxel-based analysis identified a trajectory- and discretization-dependent mobility threshold of $D_{\mathrm{crit}}^{*}$ = 3.72$\times$10$^{-7}$ cm$^{2}$/s (bootstrap 95\% CI: 2.45--5.11$\times$10$^{-7}$ cm$^{2}$/s). At this threshold, the spanning network occupied 4.2\% of voxels under the specified voxel and connectivity definitions. This result indicates that a sparse subset of high-mobility regions can form a connected cluster in the analyzed mobility field (Figure~\ref{fig:f2}(e,f)). A claim of Li$^{+}$ transport spanning the SEI should be made only after confirming that the analysis excludes metallic-atom self-diffusion and that the connected network links the physically defined SEI boundary regions.

The liquid-band Li first-shell RDF (t300x, 61 frames) shows Li--O(EC), Li--O(DMC), and Li--F(PF$_{6}$$^{-}$) competing for coordination at contact distance, with coordination numbers (CNs) of approximately 1.3, 1.3, and 1.4 (Figure~\ref{fig:f2}(c)), while the first-shell peak of carbonate-like O (CN $\approx$ 0.02) indicates only minor coordination participation of \chem{LEDC}-derived fragments.

\paragraph{Role of system scale.}
The model places multiple SEI phases, their interfaces, and adjacent organic and liquid regions within a common atomistic representation. The 43-grain mosaic permits comparisons among grain interiors, homophase boundaries, heterophase interfaces, and triple junctions, while the single-interface transport model allows local mobility and spatial connectivity to be analyzed in the same heterogeneous structure. The liquid analysis band contains approximately 688 Li ions at a reported concentration of 0.87 M, providing a basis for local diffusivity and coordination analyses without relying solely on a separate bulk-electrolyte model. Single-GPU throughput of $\approx$33 ns/day (full cell, FP32) and $\approx$25 ns/day (transport line, FP64) make the four-lane relaxation comparison plus the four-temperature Arrhenius matrix a days-long rather than months-long campaign.

\paragraph{Validation and limitations.}
Several limitations remain. No ionic conductivity is reported for the inorganic mosaic because a validated mobile-carrier population and corresponding charge-transport estimate have not been established. For the metallic Li slab, Li-atom self-diffusion does not quantify electronic conductivity, which was not calculated here. The interfaces are idealized within the modelled constituent grains, and charged-defect redistribution and explicit space-charge layers are not resolved. Finally, the Nernst--Einstein conductivity is used as a self-diffusion-based reference estimate that neglects distinct-ion cross-correlations. It is not assumed to be a strict upper bound on the true conductivity or to be directly equivalent to experimental conductivity. 

\subsection{Polycrystalline transport and simulation scale}
\label{sec:polycrystalline-li6ps5cl}

\paragraph{Scientific objective and simulation setup.}
This study examines how anion disorder and grain boundaries affect Li⁺ diffusion in polycrystalline Li₆PS₅Cl, and how grain size affects its effective conductivity. It also assesses the system sizes accessible to all-atom molecular dynamics with Symmetrix-XL. The Li⁺ flux distribution in panel (a), filled-symbol activation-energy curves in panel (b), and green conductivity points in panel (c) are calculated here with the MACE-MP small potential~\cite{mace_mp0} using Symmetrix-XL. For comparison, the open symbols in panel (b) and grey curve in panel (c) reproduce published MTP diffusion activation energies and MTP–finite-element conductivity predictions, respectively~\cite{ou2026Microstructural}. Panel (d) reports the backend performance measured in this study.

Transport simulations use LAMMPS with periodic boundary conditions. Symmetrix-XL evaluates interatomic interactions with the MACE-MP small potential. After structural relaxation and equilibration, production trajectories for Li⁺ are generated, and diffusion coefficients are extracted from the time dependence of the mean-square displacement. Arrhenius fits to diffusion coefficients at different temperatures give the activation energies. Diffusion coefficients are then converted to conductivities using the Nernst–Einstein relation to compare the effects of anion disorder, grain boundaries and grain size on transport.

The performance tests use the same pretrained MACE-MP-0a small model. Conversion to each backend preserves its architecture and learned weights. Two separately compiled LAMMPS programs integrate Symmetrix-XL and ML-IAP + cuEquivariance, respectively. Tests use four NVIDIA A100 80 GB GPUs, with one MPI rank per GPU and four ranks in total. The spatial decomposition is 2×1×2 or 1×2×2 depending on the configuration and is the same for both backends on each configuration. Calculations use FP32 precision, the NVT ensemble at 300 K, and a 0.5 fs time step. A 6 Å potential cutoff and 0.5 Å neighbor-list skin give a 6.5 Å neighbor-search range. The need to rebuild the neighbor list is checked every step. Symmetrix-XL uses streamed-edge message passing, precompiled execution kernels and a capacity configuration that permits fixed workspaces. The ML-IAP backend enables cuEquivariance fused tensor products.

Each performance test has 10 warm-up steps followed by 50 measured steps. The complete LAMMPS loop time is divided by the step and atom counts. Timing includes energy and force evaluation, MPI communication, time integration, and any neighbor-list updates within the loop. It excludes model loading, initialization, kernel compilation and warm-up. Each valid point reports the median of 10 independent runs with identical settings, with error bars spanning the minimum and maximum. All plotting statistics come from the raw logs of individual runs.

\paragraph{Observed response.}
Panel (a) shows the Li⁺ flux distribution in grain interiors and grain-boundary regions. In panel (b), both MACE-MP small and MTP give lower diffusion activation energies at intermediate disorder than at the ordered endpoints. The relative grain-boundary and bulk barriers also change with disorder, although the two models differ in their activation-energy values. MACE-MP small conductivity increases with grain size in panel (c), following the direction of the published MTP–finite-element prediction. For systems completed by both backends, Symmetrix-XL takes 0.771–0.779 μs/atom/step, compared with 7.414–9.620 μs/atom/step for ML-IAP + cuEquivariance, a 9.55–12.47-fold speedup (panel (d)). Symmetrix-XL completes short MD tests on the 25 nm polycrystalline network of 14,964,638 atoms at a median cost of 0.981 μs/atom/step. The baseline encounters memory-allocation failures for the tested 6, 10 and 25 nm configurations.

\begin{figure}[!htbp]
\centering
\includegraphics[
  width=\linewidth,
  height=0.55\textheight,
  keepaspectratio
]{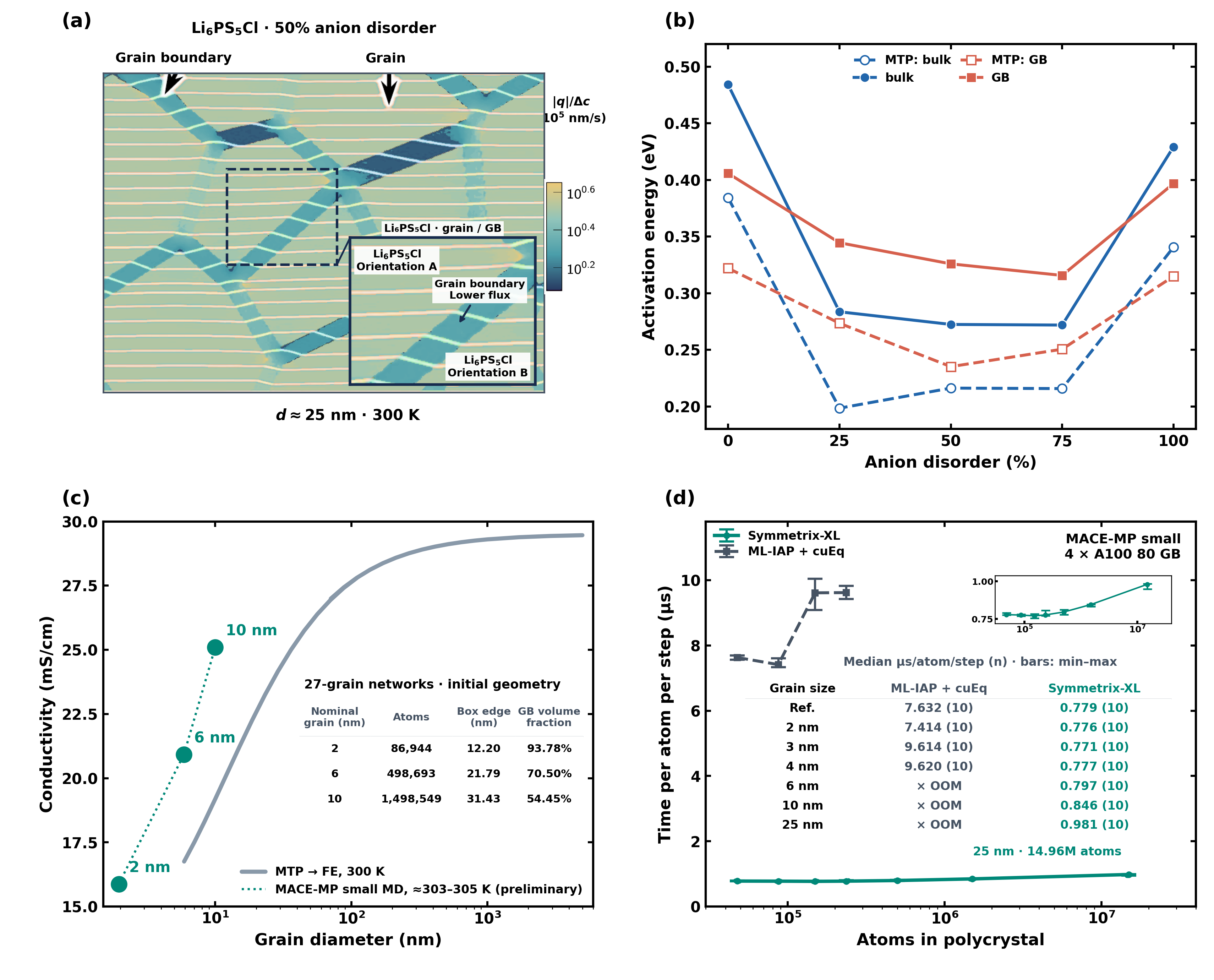}
\caption{\textbf{Transport simulations and system-size scaling in polycrystalline Li₆PS₅Cl with MACE-MP small and Symmetrix-XL.} \textbf{(a)} Normalized Li⁺ flux distribution in Li₆PS₅Cl with 50\% anion disorder at 300 K and a grain size of approximately 25 nm. The colour scale shows the local Li⁺ flux magnitude divided by the imposed concentration difference, |q|/Δc. The inset enlarges the grain interiors and adjacent grain-boundary region. \textbf{(b)} Li⁺ diffusion activation energies in the bulk and grain boundaries as a function of anion disorder. Filled symbols show MD results obtained with the MACE-MP small potential using Symmetrix-XL. Open symbols show published MTP results~\cite{ou2026Microstructural}. Blue circles and red squares denote the bulk and grain boundaries, respectively. \textbf{(c)} Effective conductivity of polycrystalline Li₆PS₅Cl as a function of grain size. Green symbols show MD results obtained with the MACE-MP small potential using Symmetrix-XL. The grey curve is the published MTP–finite-element prediction at 300 K~\cite{ou2026Microstructural}. The table lists atom counts, box lengths and grain-boundary volume fractions of the initial 27-grain networks. \textbf{(d)} Time per atom per step for the complete MD loop using the same MACE-MP small model with Symmetrix-XL or ML-IAP + cuEquivariance on four A100 80 GB GPUs. Both backends use the same model architecture and learned weights. Each run has 10 warm-up steps followed by 50 measured steps. Each valid point is the median of 10 independent runs, with error bars spanning the minimum and maximum. The inset enlarges the Symmetrix-XL timing range. × OOM marks a memory-allocation failure for ML-IAP + cuEquivariance in the corresponding configuration.
}
\label{fig:polycrystalline-li6ps5cl}
\end{figure}

\paragraph{Role of system scale.}
The MACE-MP small simulations in this study show trends similar to those reported in the published study for activation energy as a function of anion disorder and conductivity as a function of grain size. For Li₆PS₅Cl with 50\% anion disorder, the published study predicts slower diffusion at grain boundaries than in the bulk. As grains grow, the grain-boundary fraction falls and effective conductivity approaches the bulk limit, consistent with the increase in conductivity obtained in this study. Studying this size effect requires grain interiors and an interconnected grain-boundary network in the same all-atom system. In the fixed 27-grain network, increasing nominal grain size from 2 nm to 6 and 10 nm raises the atom count from 86,944 to 498,693 and 1,498,549. The initial geometric grain-boundary volume fraction falls from 93.78\% to 70.50\% and 54.45\%. Symmetrix-XL lowers the cost per atom per step and supports MD on networks of more than ten million atoms. All-atom studies can therefore cover larger grains and lower grain-boundary volume fractions to test the predicted size effect.

\paragraph{Validation and limitations.}
On the reference configuration, the two backends differ by at most 3.72×10⁻⁴ eV/Å in force components and by 8.33×10⁻⁷ eV per atom in energy. Their close numerical agreement on this configuration supports performance comparisons with the model held fixed. Quantitative comparisons of the activation energies and conductivities calculated here with published values must account for differences in potentials, temperature, microstructure and sampling convergence. The short tests in panel (d) measure execution efficiency and accessible system size. The test structures have not reached transport equilibrium, and timing error bars do not represent uncertainty in transport coefficients. Networks of more than ten million atoms can now be run, but reliable effective conductivity at that scale still requires sufficient equilibration and sufficiently long production trajectories.

\FloatBarrier

\subsection{Structural and Diffusive Evolution of the Liquid Au--Sn Alloy/Solid Au Interface}
\label{sec:ausn-au-diffusion}

\begin{figure}[!htbp]
\centering
\includegraphics[page=1,width=\textwidth]{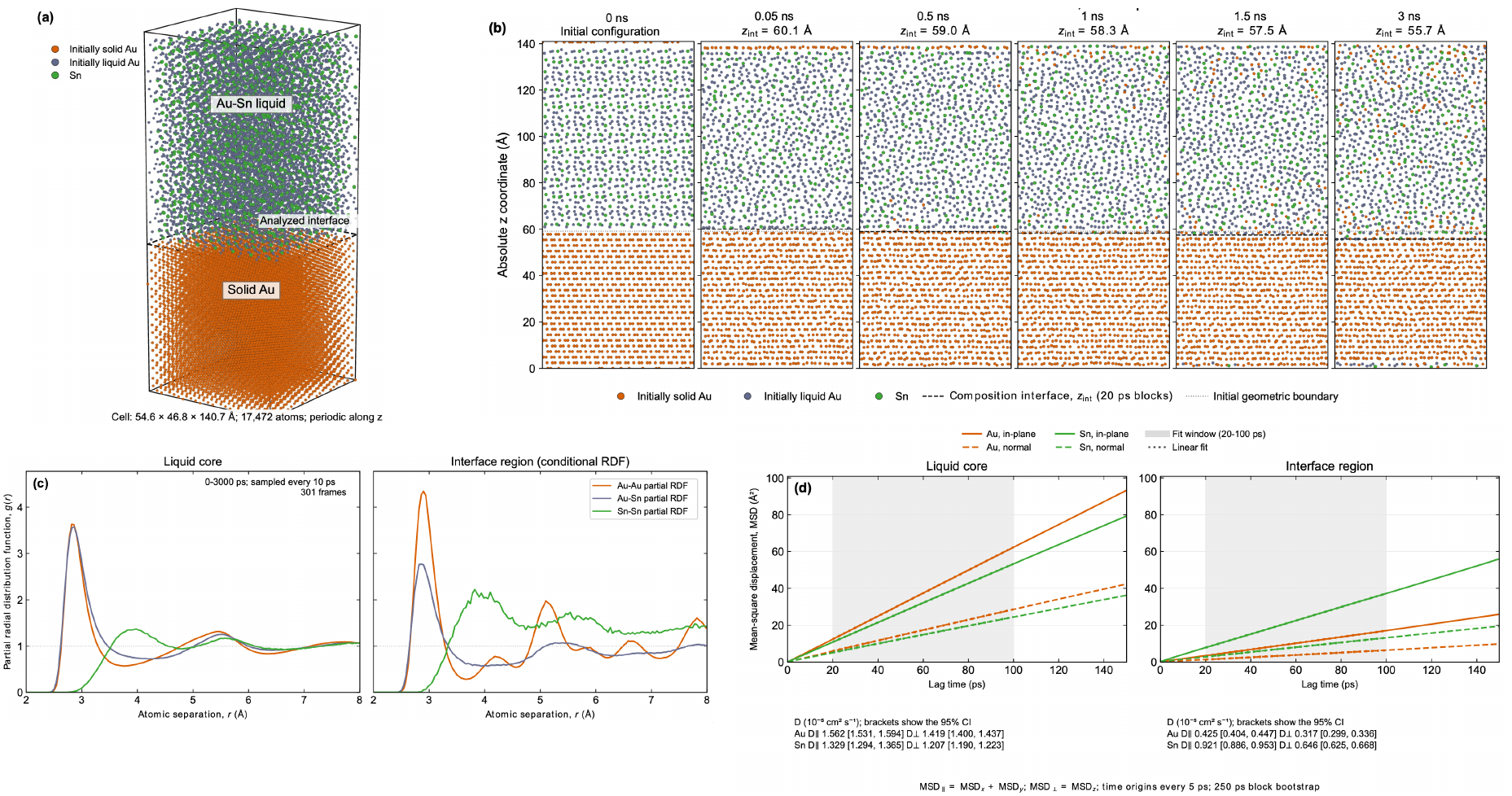}
\caption{Structural and diffusive evolution of the liquid Au--Sn
alloy/solid Au interface. (a) Initial interfacial model containing 17,472
atoms; colors indicate the initial origins of the atoms. (b) Structural
evolution at 603 K from 0 to 3000 ps; the dashed line denotes the compositional
interface, $z_{\mathrm{int}}$, determined from Sn concentration profiles in
20 ps blocks, and the dotted line denotes the initial geometric interface.
(c) Liquid-core partial RDFs and conditional interfacial RDFs sampled every
10 ps over 301 frames. (d) In-plane and normal MSDs of Au and Sn in the liquid
core and interfacial region; the shaded region indicates the 20--100 ps fitting
interval, and the diffusion coefficients and their 95\% confidence intervals
are listed below the panels.}
\label{fig:ausn-au-interface}
\end{figure}

\paragraph{Scientific objective and simulation setup.}
This study examines dissolution and interdiffusion between a liquid Au--Sn alloy and an Au(111) substrate at 603 K. The 17,472-atom model contains 8,400 substrate Au atoms and a liquid region of 7,224 Au and 1,848 Sn atoms in a $54.626\times46.823\times140.685\ \text{\AA}^3$ cell with a $2531.5\ \text{\AA}^2$ lateral interface and a $2.700\ \text{\AA}$ initial gap. Periodicity along $z$ creates two interfaces; the internal interface in Figure~\ref{fig:ausn-au-interface}a is analyzed. The 3 ns trajectory comprised three stages: 0--200, 200--1451, and 1451--3000 ps. Restarts recovered the cell, coordinates, and momenta without velocity reinitialization; duplicate boundary frames were removed, and energy, temperature, and cell parameters remained continuous. All stages shared float32 precision, a 1 fs timestep, 603 K, nominal 1 atm, a $z$-only NPT workflow, and 100 fs thermostat/barostat coupling times. They used a Symmetrix potential converted from the same OMat24-trained MACE-OMAT medium checkpoint~\cite{barrosLuque2026OMat24}, with the standard backend for the first two stages and the optimized backend for the third. Frames were saved every 1 ps (3001 total), and Au colors denote initial origin rather than instantaneous structure (Figure~\ref{fig:ausn-au-interface}b).

\paragraph{Observed response.}
Substrate-origin Au progressively appears on the liquid-alloy side while the central Au lattice remains ordered. The 20 ps block-averaged compositional interface, $z_{\mathrm{int}}$, moves from $60.1\ \text{\AA}$ at 50 ps to $55.7\ \text{\AA}$ at 3000 ps, a retreat of approximately $4.35\ \text{\AA}$ toward the solid (Figure~\ref{fig:ausn-au-interface}b). Requiring atoms to cross at least $3\ \text{\AA}$ beyond the dynamic interface and remain on the destination side for at least 80\% of the preceding 20 ps window identifies 174 persistent substrate-Au-to-liquid events, but no persistent liquid-Au-to-solid or Sn-to-solid events. Conditional RDFs sampled every 10 ps over 301 frames give liquid-core Au--Au, Au--Sn, and Sn--Sn first peaks at $2.86$, $2.86$, and $3.94\ \text{\AA}$, compared with $2.90$, $2.86$, and $3.86\ \text{\AA}$ at the interface. The corresponding first-shell coordination numbers change from 8.46, 2.92, and 3.20 to 10.58, 0.96, and 2.34, indicating Au enrichment without establishing a new crystalline compound (Figure~\ref{fig:ausn-au-interface}c). Fits to the 20--100 ps MSD regime give liquid-core $(D_{\parallel}/D_{\perp})$ values of $(1.562/1.419)\times10^{-5}\ \mathrm{cm}^2\,\mathrm{s}^{-1}$ for Au and $(1.329/1.207)\times10^{-5}\ \mathrm{cm}^2\,\mathrm{s}^{-1}$ for Sn; the interfacial values decrease to $(0.425/0.317)\times10^{-5}$ and $(0.921/0.646)\times10^{-5}\ \mathrm{cm}^2\,\mathrm{s}^{-1}$, respectively. Bootstrap confidence intervals are reported in Figure~\ref{fig:ausn-au-interface}d, which shows slower interfacial motion and stronger suppression in the normal direction.

\paragraph{Role of system scale.}
The approximately $55\times47\ \text{\AA}^2$ interface, 17,472 atoms, and continuous 3 ns trajectory jointly resolve lateral heterogeneity, slow interface retreat, and discrete persistent crossings; smaller or shorter calculations would be more sensitive to periodic images, composition fluctuations, and transient thermal crossings. Each liquid-core RDF frame contains on average approximately 5781 Au and 1381 Sn centers, while the interfacial RDF contains approximately 1176 Au and 98 Sn centers, providing smooth element-resolved statistics (Figure~\ref{fig:ausn-au-interface}c). Multiple time origins and twelve 250 ps blocks support confidence intervals and reveal $D_{\parallel}/D_{\perp}$ ratios of approximately 1.10 in the liquid and 1.34--1.43 at the interface (Figure~\ref{fig:ausn-au-interface}d). The optimized single-A100 backend reaches approximately $906\ \mathrm{ps/day}$, making this $10^4$-atom, nanosecond-scale calculation accessible within several GPU-days. Because no controlled size comparison was performed, these results demonstrate improved spatial and statistical sampling but not strict size convergence.

\paragraph{Validation and limitations.}
Interface location and diffusion were evaluated with custom Python scripts. Error-function fits to 20 ps Sn profiles used $0.5\ \text{\AA}$ bins and $0.8\ \text{\AA}$ smoothing; 295 of 300 time-block--interface fits met $R^2\geq0.80$, and adjacent-block displacements were at most $3\ \text{\AA}$. Atomic colors encode initial origin, not CNA, PTM, or $q_6$ classifications, and RDFs do not assign atom-wise solid or liquid states. Diffusion coefficients were obtained from multiple-time-origin Einstein MSDs with periodic unwrapping, 5 ps origin spacing, and drift correction using 4368 solid-core Au atoms; the 20--100 ps fits gave $R^2>0.9998$, with scaling-exponent checks and 250 ps block bootstrap uncertainties (Figure~\ref{fig:ausn-au-interface}d). Trajectory-continuity, unwrapping, fit-quality, and internal-consistency checks support these procedures, but no independent implementation or reference-system benchmark has been performed; validation is therefore internal, not external. The block-averaged $z_{\mathrm{int}}$ and interfacial RDFs and diffusivities are regional effective measures. Finally, one temperature, one initial configuration, and one 3 ns trajectory establish net dissolution only under the stated conditions and event criterion, not a general mechanism; testing generality requires temperature series, independent trajectories, threshold tests, and targeted DFT, while the confidence intervals exclude potential-model error.

\FloatBarrier

\section{Discussion and limitations}
\label{sec:discussion}

Symmetrix-XL adds three execution mechanisms on top of Symmetrix \cite{symmetrix_original}: streamed-edge execution removes graph-wide expanded edge state, model-specialized execution compiles the fixed operator structure of a trained checkpoint, and tiled execution bounds workspace when graph-wide state limits capacity.
Together, they improve throughput and single-accelerator scale-up without changing the model's learned weights or architectural contract, although the end-to-end gains describe their combined implementation rather than any mechanism in isolation.

This route complements architecture optimization and distillation, which change the deployed model \cite{nequix,uma,e2former,dpa4c,sevennet_nano}, and framework-level approaches such as compiled JAX deployment \cite{chemtrain_deploy}.
These strategies address different parts of the accuracy--cost trade-off and may be combined; our results show that substantial optimization remains possible for existing models \emph{without} modifying their architecture.

The results motivate co-design of model architecture and inference implementation.
Streaming benefits architectures whose edge-local intermediates can be consumed during receiver aggregation, while fixed sparse structure enables specialization.
These execution patterns are not unique to MACE: other equivariant models may benefit from the same streaming, specialization, and tiling principles when their operators admit compatible schedules, although each architecture requires its own implementation and numerical qualification.
Conversely, numerous species-, channel-, or edge-conditioned radial functions enlarge the spline-table working set and may reduce cache reuse.
Appendix~\ref{app:mpi-one} shows increasing per-atom cost as the number of species and spline tables grows, although attributing this trend to cache behavior remains an interpretation rather than a hardware-counter result.

The scale-up from tiled streaming is particularly relevant to accelerators with smaller memory capacities.
From the measured 32~GB tiled-streaming boundary, we estimate capacities of approximately 1.7 and 2.9 million atoms for 16 and 24~GB devices, respectively, for the same SrTiO$_3$ system, MACE-OMAT-0 checkpoint, cutoff, skin, and precision.
This indicates that million-atom simulations should be feasible on such devices for comparable workloads.

We note that absolute performance is platform dependent and sensitive to the software stack, precision, checkpoint, neighbor graph, and timing boundary.
Our matched comparisons use the same local hardware and complete ASE-call or LAMMPS-step boundaries, but the timings should not be assumed to transfer unchanged across platforms, and short capacity probes do not establish sustained molecular-dynamics performance.
Efficiency comparisons should therefore report their deployment path and use competitive optimized implementations with matched measurement boundaries, particularly when making accuracy--cost claims.

\section{Conclusion}
\label{sec:conclusion}

Streamed-edge execution, model-specialized code generation, and tiled execution substantially improve the throughput and single-accelerator capacity of trained MACE models without retraining or changing their learned weights and architectural contract.
The optimized path reaches 11.24 million atoms on one A100 80~GB, composes with spatially decomposed multi-GPU execution, and reproduces reference MACE predictions and downstream Matbench Discovery results within the reported tolerances.
The case studies demonstrate the resulting capacity in large solid-state and reactive molecular simulations.
More broadly, the execution patterns developed here may be applicable to other equivariant models whose fixed operators and edge-local intermediates admit specialization, streaming, and tiled evaluation.
These results establish dedicated inference backends as a complementary route to scaling foundation models and motivate future model architectures designed with efficient post-training execution in mind.

\section*{Code Availability}

The Symmetrix-XL source code is available at \url{https://github.com/bonan-group/symmetrix-xl}.

\section*{Acknowledgements}

The original Symmetrix code was developed by William C. Witt and is available from its public repository \cite{symmetrix_original}.
We thank Alex M. Ganose and Seán R. Kavanagh for fruitful discussions.
This work was supported by the Fundamental Research Funds for the Central Universities through the Beijing Institute of Technology academic start-up grant RCPT-6120230091.
B. Zhu acknowledges support from the National Natural Science Foundation of China (Grant No. 12404068).
This work was also supported by Zhongguancun Academy (Grant No. C20250609).

\bibliographystyle{unsrtnat}
\bibliography{references}

\clearpage

\appendix

{\Large\bfseries Appendix\par}
\vspace{1em}

\section{Additional Fidelity Verification}
\label{app:additional-fidelity}

\begin{figure*}[htbp]
    \centering
    \includegraphics[width=\textwidth]{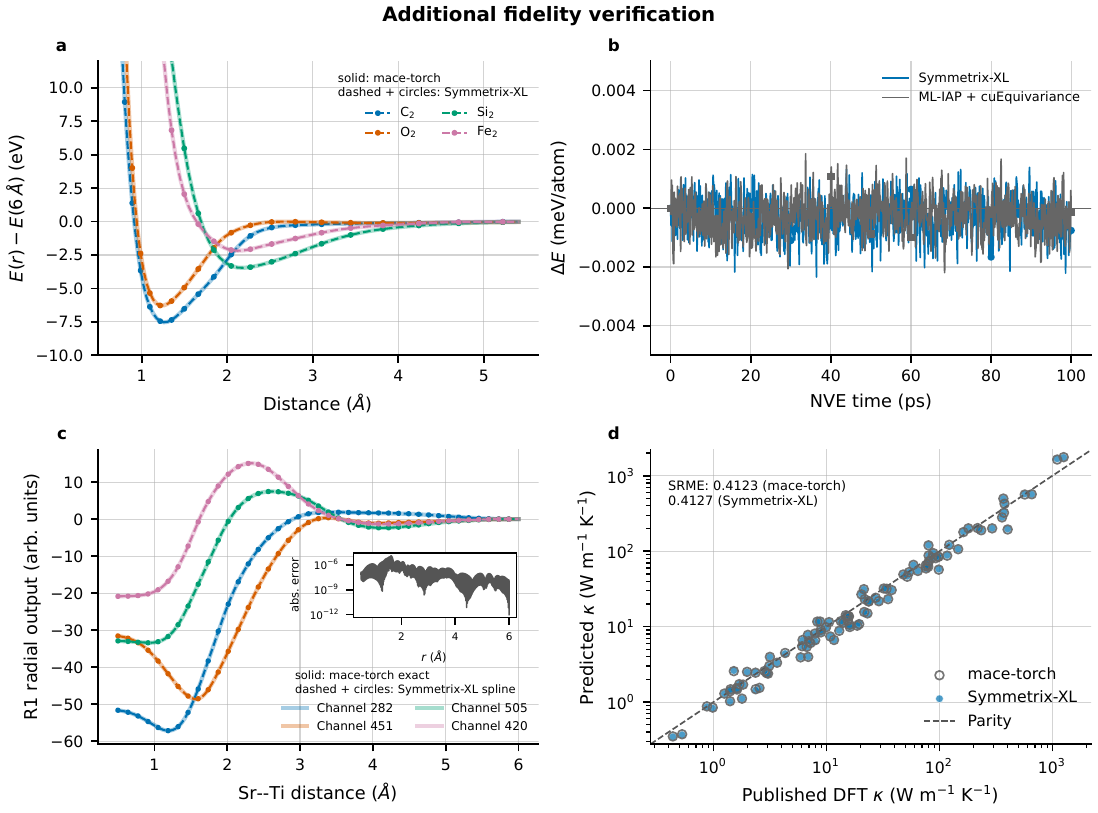}
    \caption{Additional fidelity verification.
    \textbf{(a)} Selected homonuclear MACE-MPA-0 medium diatomic curves from
    Symmetrix-XL and the published mace-torch evaluation \cite{mace_mp0}. Energies
    are referenced to the value at 6~\AA{} for each element. Across all 87
    supported element pairs and 119 distances per curve, the mean absolute
    energy difference is $5.67\times10^{-6}$~eV/dimer.
    \textbf{(b)} Change in total energy during 100~ps of NVE dynamics for the
    5,000-atom SrTiO$_3$ system after 20~ps of NVT equilibration. Both
    implementations use MACE-OMAT-0 medium, float32, and a 1~fs timestep. The
    fitted drifts are $-8.70\times10^{-10}$ and
    $+3.66\times10^{-10}$~eV/atom/ps for Symmetrix-XL and
    ML-IAP~+~cuEquivariance, respectively; the small fit $R^2$ values indicate
    bounded fluctuations rather than resolved monotonic trends.
    \textbf{(c)} Four representative Sr--Ti R1 radial-network output channels
    from the extracted MACE-MPA-0 medium checkpoint and their 256-node runtime
    splines over 0.5--6.0~\AA{}. Solid curves show the exact mace-torch radial
    outputs, while dashed curves with circular markers show the corresponding
    Symmetrix-XL splines. The inset shows the largest absolute
    interpolation error across the displayed channels. The ordinate is an
    internal learned feature in arbitrary units, not an energy or force.
    \textbf{(d)} Scalar 300~K lattice thermal conductivities for the 103
    structures in the published MACE-MPA-0 protocol \cite{pota_thermal}.
    Published DFT values are plotted against both the published mace-torch and
    Symmetrix-XL predictions. The annotations give the corresponding symmetric
    relative mean errors (SRME).}
    \label{fig:additional-fidelity}
\end{figure*}

Figure~\ref{fig:additional-fidelity} collects complementary checks spanning
model outputs, molecular-dynamics stability, the spline approximation used by
the evaluator, and a downstream phonon workflow.  These checks test different
parts of the execution path and are not interchangeable: the diatomic curves
measure agreement with published mace-torch results, the thermal-conductivity
results compare both implementations with DFT references, the NVE trajectory
tests conservation of a physical invariant, and the radial comparison measures
interpolation of internal learned features.

\section{Radial spline construction}
\label{app:splines}

Symmetrix projects eligible radial networks onto cubic Hermite splines during model preparation.
For nodal values \(y_i\), nodal slopes \(m_i\), interval width \(h\), and \(\tau=(r-r_i)/h\), the stored polynomial is
\[
S_i(r)=(2\tau^3-3\tau^2+1)y_i+(\tau^3-2\tau^2+\tau)hm_i+(-2\tau^3+3\tau^2)y_{i+1}+(\tau^3-\tau^2)hm_{i+1}.
\]
The radial derivative is the analytic derivative of the same polynomial, so forces and stress are consistent derivatives of the approximated energy.
The learned linear projection can be absorbed into the table because spline construction is linear in the tabulated values.

Each checkpoint is compared with its exact radial basis, cutoff, distance transform, and radial network on a combined linear and logarithmic radius grid.
The gate checks agreement at spline nodes, interpolation error between nodes, radial derivatives over the declared physical range, and end-to-end energy, force, and stress deviations.
Unsupported radial families are rejected rather than silently approximated.
The MACE-OMAT-0-medium qualification passes at 256 spline nodes, with worst internal-feature value and derivative interpolation errors of \result{$5.62\times10^{-5}$} and \result{$7.23\times10^{-3}$}, respectively, over the declared physical range.
These quantities are expressed in the learned radial network's internal output units rather than in units of energy, force, or stress; physical fidelity is assessed separately through the end-to-end comparisons above.

\section{Element-count scaling benchmark}
\label{app:mpi-one}

A simulation cell containing \num{4800} atoms was used without OpenMP threading, and the MPI rank count was varied among 1, 12, and 48.
The element-count series contains 1, 5, and 20 species.
All runs were performed on a node equipped with dual AMD EPYC 9654 processors and 768~GB of DDR5-5600 memory.

\begin{figure}[!htbp]
    \centering
    \includegraphics[width=\textwidth]{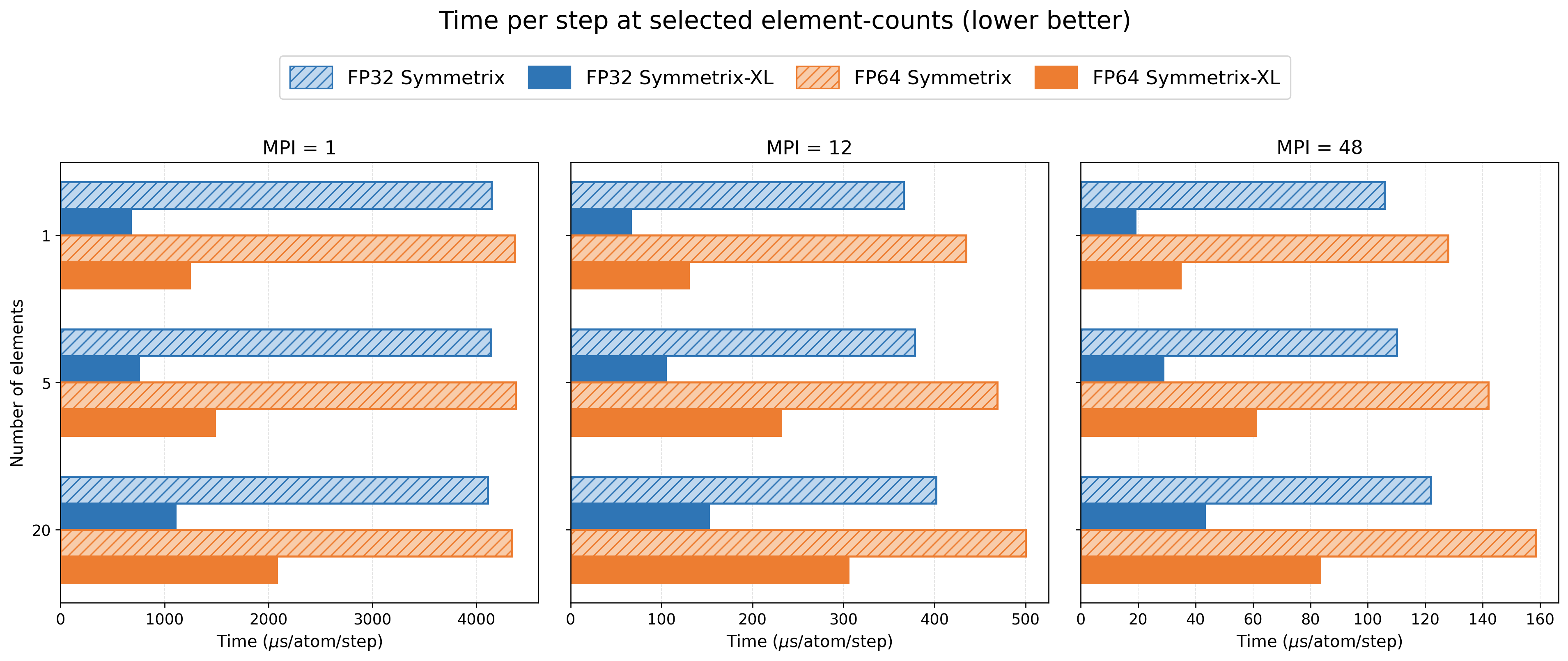}
    \caption{MPI-only element scaling for the \num{4800}-atom benchmark with
    OpenMP fixed to one thread.  The combined figure contains the 1-, 
    5-, and 20-element systems across the tested MPI rank counts, with
    matched implementation and precision comparisons.  Bars report time in
    \si{\micro\second}/atom/step.}
    \label{fig:mpi-one-elements}
\end{figure}

Each case uses five warm-up steps followed by 20 measured NPT steps. 
The reported observable is the final measured time per atom per step; lower values
indicate higher performance. 
The comparison includes FP32 and FP64 for both
the original Symmetrix implementation and Symmetrix-XL.

Figure~\ref{fig:mpi-one-elements} shows the three element-count groups.  
Symmetrix-XL is approximately five times faster than the original Symmetrix implementation for the matched cases. 
However, the per-atom cost increases as the number of element species grows from 1 to 20, whereas the original Symmetrix only shows this trend at high MPI rank counts.

For Symmetrix-XL, FP32 gives an approximately twofold speed advantage
over FP64.  
This behavior is consistent with the SIMD execution path: at a fixed
vector width, FP32 processes twice as many scalar lanes as FP64, provided that the kernels remain compute-bound and use the corresponding AVX2/AVX-512 instructions.  
The original Symmetrix implementation shows little separation between
FP32 and FP64 under the same benchmark,
indicating that it does not benefit from SIMD in this configuration.

\end{document}